\documentclass[twocolumn]{aastex6}

\usepackage{graphicx}
\usepackage{amsmath}
\usepackage{amssymb}

\newcommand{\mbf}[1]{\mbox{\boldmath $#1$}}
\newcommand{\mbfs}[1]{\mbox{\scriptsize\boldmath $#1$}}

\newcommand{\Eqn}[1]{Equation~(\ref{eqn:#1})}
\newcommand{\Eqns}[3]{Equations~(\ref{eqn:#1}) #2~(\ref{eqn:#3})}

\newcommand{\Sec}[1]{Section~\ref{sec:#1}}
\newcommand{\Secs}[3]{Sections~\ref{sec:#1} #2~\ref{sec:#3}} 

\newcommand{\Fig}[1]{Figure~\ref{fig:#1}}

\newcommand{\Tab}[1]{Table~\ref{tab:#1}}

\newcommand{\App}[1]{Appendix~\ref{app:#1}}

\newcommand{\rankfour}{\ensuremath{\mathbb{C}^2_{2;2}}}
\newcommand{\ranktwo}{\ensuremath{\mathbb{C}^4_{1;1}}}

\newcommand{\irow}{\mu} \newcommand{\icol}{\nu}
\newcommand{\jrow}{\kappa} \newcommand{\jcol}{\lambda}
\newcommand{\srow}{j} \newcommand{\scol}{k}

\newcommand{\trace}{{\rm Tr}}
\newcommand{\real}{{\rm Re}}
\newcommand{\imag}{{\rm Im}}

\newcommand{\tr}[1]{\trace\ensuremath{ \left[ {#1} \right] }}
\newcommand{\re}[1]{\real\ensuremath{ \left[ {#1} \right] }}
\newcommand{\im}[1]{\imag\ensuremath{ \left[ {#1} \right] }}

\newcommand{\mean}[1]{\ensuremath{ \langle #1 \rangle }}

\newcommand{\Linner}[2]{\ensuremath{ { {#1} \cdot {#2} } } }

\newcommand{\bilinear}[2]{\ensuremath{\left({#1},{#2}\right)}}

\newcommand{\outerBilinear}[2]{\ensuremath{{#1}\otimes{#2}}}
\newcommand{\stimes}{\ensuremath{\tilde{\otimes}}}
\newcommand{\spinorBilinear}[2]{\ensuremath{{#1}\,\stimes\,{#2}}}

\newcommand{\outerDiff}[2]{\ensuremath{({#1}-{#2})^{\otimes 2}}}
\newcommand{\outerSymm}[2]{\ensuremath{{#1}\,\tilde{\odot}\,{#2}}}

\newcommand{\outerMueller}[2]{\ensuremath{{\bf M}_\otimes\bilinear{#1}{#2}}}
\newcommand{\spinorMueller}[2]{\ensuremath{{\bf M}_{\stimes}\bilinear{#1}{#2}}}

\newcommand{\exterior}[2]{\ensuremath{{#1}\wedge{#2}}}
\newcommand{\dc}[2]{\ensuremath{{#1}\,\mbf{:}\,{#2}}}

\newcommand{\pauli}[1]{\ensuremath{\mbf{\sigma}_{#1}}}

\newcommand{\inv}[1]{\ensuremath{ {#1}^2 }}
\newcommand{\norm}[1]{\ensuremath{ \|{#1}\| }}

\newcommand{\element}[3]{\ensuremath{ \left\{ {#1} \right\}_{#2}^{#3} }}
\newcommand{\Celement}[1]{\element{#1}{\irow}{\icol}}

\newcommand{\instI}{\ensuremath{\xi}}

\shorttitle   {Disjoint, Superposed, and Composite Samples}
\shortauthors {van Straten \& Tiburzi}

\received{}

\begin{document}

\title{ The Statistics of Radio Astronomical Polarimetry: \\
	Disjoint, Superposed, and Composite Samples }

\author{W. van Straten}
\affil{Centre for Astrophysics and Supercomputing,
	Swinburne University of Technology,
	Hawthorn, VIC 3122, Australia}
\affil{ARC Centre of Excellence for All-sky Astrophysics (CAASTRO)}
\affil{Institute for Radio Astronomy \& Space Research,
Auckland University of Technology,
Private Bag 92006,
Auckland 1142,
New Zealand}
\email{willem.van.straten@aut.ac.nz}

\author{C. Tiburzi}
\affil{Max-Planck-Institut f\"{u}r Radioastronomie,
       Auf dem H\"{u}gel 69, 53121 Bonn, Germany}
\affil{Universit\"{a}t Bielefeld, Fakult\"{a}t f\"{u}r Physik,
       Universit\"{a}tsstr. 25, D-33615 Bielefeld, Germany}

\begin{abstract}

A statistical framework is presented for the study of the orthogonally
polarized modes of radio pulsar emission via the covariances between
the Stokes parameters.  To accommodate the typically heavy-tailed
distributions of single-pulse radio flux density, the fourth-order
joint cumulants of the electric field are used to describe the
superposition of modes with arbitrary probability distributions.  The
framework is used to consider the distinction between superposed and
disjoint modes with particular attention to the effects of integration
over finite samples.  If the interval over which the polarization
state is estimated is longer than the timescale for switching between
two or more disjoint modes of emission, then the modes are unresolved
by the instrument.  The resulting composite sample mean exhibits
properties that have been attributed to mode superposition, such as
depolarization.  Because the distinction between disjoint modes and a
composite sample of unresolved disjoint modes depends on the temporal
resolution of the observing instrumentation, the arguments in favour
of superposed modes of pulsar emission are revisited and observational
evidence for disjoint modes is described.  In principle, the
four-dimensional covariance matrix that describes the distribution of
sample mean Stokes parameters can be used to distinguish between
disjoint modes, superposed modes, and a composite sample of unresolved
disjoint modes.  More comprehensive and conclusive interpretation of
the covariance matrix requires more detailed consideration of various
relevant phenomena, including temporally correlated subpulse
modulation (e.g.\ jitter), statistical dependence between modes
(e.g.\ covariant intensities and partial coherence), and multipath
propagation effects (e.g.\ scintillation and scattering).
Unpublished supplementary material is appended after the bibliography.

\end{abstract}

\keywords{methods: data analysis --- methods: statistical 
--- polarization --- pulsars: general --- techniques: polarimetric}


\section {Introduction}

The higher-order statistics of electromagnetic radiation have long
been used to discover and study radio pulsars.
The fluctuation power spectrum is employed to detect periodic
signals in pulsar survey data \citep[e.g.][]{bc69,rem02};
the longitude-resolved modulation index is used to study the
radio pulsar emission mechanism
\citep[e.g.][]{tmh75,kd83,jg03,wse07};
correlated and periodic structure in the variability of pulsar
signals, such as drifting subpulses, are detected and studied using
longitude resolved and two-dimensional fluctuation spectra
\cite[e.g.][]{bac70,es02};
and the secondary dynamic spectrum reveals information about the
turbulent structure of the ionized interstellar medium along the
pulsar line of sight \cite[e.g.][]{smc+01,wmsz04,crsc06}.
All of the above quantities involve fourth-order moments of the
electric field that characterize variability in the flux density of
the pulsar signal.

The fourth moments of the electric field have also been used to study
variability in the polarization of pulsar radiation.
%
%
In many pulsars, over limited ranges of pulsar longitude, the emission
is observed to switch between one of two orthogonally polarized
states, or modes \citep[e.g.][]{em69,thhm71,mth75,br80,scr+84}.
%
%
The degree of polarization of the modes and the time scales for mode
switching (e.g.\ microstructure and subpulse polarization
fluctuations) have been inferred from the autocorrelation functions of
the Stokes parameters \citep[e.g.][]{cor76,ch77,cbh+04}.
%
In several cases, the observed modes are not perfectly orthogonally
polarized \citep[e.g.][]{br80,scr+84,gss91,mck03a}
%
and surprising annular distributions of single-pulse polarization states
on the surface of the Poincar\'{e} sphere have also been observed,
which may be indicative of stochastic generalized Faraday rotation in
the pulsar magnetosphere \citep{es04}.

\citet{crb78}, hereafter CRB, first proposed using the (non-central)
second moments of the linearly polarized and total intensities to
determine if the modes are \emph{disjoint}, such that only one mode is
observed at a given instant, or if the signal is an incoherent
\emph{superposition} of the two modes.
CRB also argued that the high degree of correlation between the
handedness of circular polarization and the position angle of the
linearly polarized flux is further evidence of the mutual exclusivity
of the modes.
However, this correlation is more simply interpreted as evidence for
the ellipticity of the modes \citep[e.g.][]{ht81,am82,kjk03}.

Over the decades following CRB, several studies considered the distinction
between disjoint and superposed modes \citep[e.g.][]{scr+84,ms98}, and
a wide variety of statistical approaches to the study of orthogonally
polarized modes have been developed
\citep[e.g.][]{ms00,mck02,mck03b,mck06}.
\citet{es04} and \citet{mck04} independently introduced techniques
based on principal component analysis of the $3\times3$ covariance
matrix of the Stokes polarization vector with the respective goals of
characterizing the non-orthogonality of the modes and testing the
hypothesis that the excess broadening of position angle histograms
\citep[first noted by][]{scr+84} is due to additional randomly
polarized radiation.
Influenced by these studies, \citet{van09} analyzed the $4\times4$
covariance matrix of the Stokes parameters and demonstrated that
additional randomly polarized radiation is not required.  Rather, the
observed variance of the polarized flux is consistent with the self
noise intrinsic to the bright pulsars on which single-pulse studies
typically focus.

The statistical analysis presented by \citet{van09} is valid only when
the components of the electric field vector are normally distributed.
However, pulsars typically exhibit heavy-tailed (e.g.\ power-law and
log-normal) distributions of longitude-resolved single-pulse flux
density \citep[e.g.][]{cjd03a,ovb+14}.
Therefore, one of the main aims of this paper is to extend (and
correct) the statistical framework of \citet{van09} so that it can be
applied to non-normal distributions.

As in \citet{van09}, the framework presented in this paper 
differentiates between the statistics of the \emph{instantaneous}
Stokes parameters, which are derived from a single instance of the
electric field, and the \emph{sample mean} Stokes parameters, which
are computed by averaging over a finite sample of instances of the
electric field.
The instantaneous Stokes parameters, also known as the
\emph{unaveraged} or \emph{unsmoothed} Stokes parameters, are
typically encountered only when studying the autocorrelation functions
of the Stokes parameters as a function of lag
\citep[e.g.][]{cor76,ch77,cbh+04} or when presenting the polarization
of giant pulses at the highest available time resolution
\cite[e.g.][]{hcr70,cstt96,hkwe03}.\footnote{See \citet{van09} for a
  detailed discussion of the fundamental limitations and pitfalls
  associated with studying giant pulse polarization via the
  instantaneous Stokes parameters, which on their own have no
  statistically significant physical meaning.}
The sample mean Stokes parameters are more commonly encountered in the
study of single pulses from radio pulsars, where the finite samples
are defined by the evenly spaced intervals of pulsar longitude, also
known as \emph{phase bins}, into which the signal is divided.

The distinction between instantaneous and sample mean statistics
reaffirms the fundamental importance of instrumental resolution when 
studying orthogonally polarized modes.
Whereas instantaneous Stokes parameters admit only disjoint or
superposed modes, the sample mean Stokes parameters may include
statistical samples that are composed of a union of sub-samples drawn
from mutually exclusive populations.
In other words, after integration, disjoint modes may be unresolved by
the instrument.

Observational evidence of disjoint modes is presented by \cite{ch77},
who find that micropulse structures on the shortest time scales are
more polarized than the longer time scale subpulse structures in which
they are embedded.  They also observe that transitions between
orthogonally polarized modes preferentially occur on the edges of
micropulses and conclude that the variability of micropulse
polarization depolarizes the signal when it is smoothed on a time
scale greater than the characteristic width of the micropulse
structures.
That is, the signal is depolarized when the disjoint modes are unresolved.
As discussed in more detail in \App{gxv+99}, similar evidence of
unresolved disjoint modes is presented by \citet{gxv+99}, who
demonstrate that the degree of polarization of single-pulse
observations is higher when the integration interval is shorter; that
is, at higher time resolution, the disjoint modes are better resolved.

Integration over a sample that is composed of mutually exclusive and
orthogonally polarized sub-samples depolarizes the resulting sample
mean Stokes parameters. This provides an alternative explanation for
depolarization, which is more commonly interpreted as evidence of mode
superposition \citep[e.g.][]{scr+84,ms98,krj+11}.
Because the degree of polarization, and therefore the previously
proposed distinction between disjoint and superposed modes, depends
on the temporal resolution of the instrument used to record the
experimental data, the observational evidence that has been presented
in support of superposed modes should be revisited.


Toward this end, and motivated by the additional insights that can be
gained through analysis of the four-dimensional fourth-order moments
of the electric field, the primary aim of this paper is to further
develop a statistical framework for the study of orthogonally
polarized modes via the covariances between the Stokes parameters.
To date, studies of variability in the polarized emission from pulsars
have focused on sources bright enough to be detected on every
rotation.
In contrast, the approach described in this paper does not require
detection of individual pulses and can be applied to average pulse
profiles of arbitrary integration length.
Therefore, the proposed framework can be employed to study sources
that are either insufficiently bright to be clearly detected in
single-pulse data or for which recording and offline analysis of such
data is prohibitively expensive, which are some of the reasons why
single-pulse studies of millisecond pulsar polarization are relatively
rare and recent \citep[e.g.][]{ovb+14,lkl+15}.
Experiments based on statistical interpretation of moments should be
less influenced by the idiosyncrasies of the relatively few brightest
(and more slowly spinning) sources and thereby have the potential to
enable more statistically significant conclusions about the entire
pulsar population.

Following a review of fourth-order statistics in \Sec{review}, some
relevant linear algebra and the covariances between the instantaneous
Stokes parameters are presented in \Sec{Stokes_statistics}.
In \Sec{modes}, the covariances between the sample mean Stokes
parameters are derived for the following three
distinct and idealized combinations of statistical samples.
\begin{enumerate}
\item \emph{Disjoint samples}: Every instance of the electric field vector in
  a given sample is exclusively from only one of the two modes of emission;
  i.e.\ the disjoint modes are resolved by the instrument.
\item \emph{Superposed samples}: Each instance of the electric field vector
  is an incoherent sum of the electric fields from the two modes.
\item \emph{Composite samples}: Each sample is a union of sub-samples of instances
  of the electric field vector from each of the two modes of emission;
  i.e.\ the disjoint modes are unresolved by the instrument.
\end{enumerate}
In \Sec{interpretation}, the results of this analysis are interpreted
with attention to the impact of various physical phenomena, including
amplitude modulation, partial mode coherence, interstellar
scintillation, and superposition of signal and noise.
The mathematical equations presented throughout this paper are
verified using a Monte Carlo simulation that is described in the
Appendix and the results of this study are summarized and discussed in
\Sec{discussion}.


\subsection{Review of Fourth-Order Statistics}
\label{sec:review}

This section reviews definitions and concepts that are relevant to the
\emph{instantaneous} Stokes parameters, beginning with the
fourth-order statistics of complex-valued random scalar variables.
One of the central mathematical results of this paper is derived using
the cumulants of a probability distribution, which are closely related
to its moments \cite[e.g.][]{kso87}.  As
demonstrated in the following sections, the cumulants have an
important property that is not shared by the moments of a
distribution. For any two statistically independent random variables,
$x$ and $y$, the cumulants of the sum $z=x+y$ are equal to the sums of
the cumulants of $x$ and $y$.
The moments and cumulants of a distribution are equal only up to third
order and, to highlight the fundamental differences between them, the
following section begins with a simple demonstration that the fourth
moment of a sum of two random variables is not equal to the sum of
their fourth moments.
A definition of the cumulants is then presented using a derivation
that explains the origin of their additive property, which
is later exploited to study the fourth-order statistics of superposed
electromagnetic waves.


\subsubsection{Complex-valued Random Scalars}
\label{sec:complex_statistics}

The following example illustrates the utility of the higher order
cumulants of a distribution when studying the superposition of
electromagnetic waves.
%
%
Consider a complex-valued analytic signal $z$ associated with a
real-valued random variable \cite[e.g.][]{bra86b}.
Such a signal is an example of a {\it complex circular} scalar
field, defined as having statistically independent and 
identically distributed real and imaginary components, $\re{z}$ and
$\im{z}$ \citep[e.g.][]{pic94}.
If $\re{z}$ and $\im{z}$ are also normally distributed, then the
instantaneous intensity of this scalar field, $\instI=z^*z$, where $z^*$ is
the complex conjugate of $z$, is distributed as $\chi^2$ with two
degrees of freedom (the exponential distribution).
Therefore the standard deviation of $\instI$ is equal to its mean;
i.e.\ $\varsigma_\instI = \mean{\instI}$, where the angular brackets denote the
expectation value, or population mean.
(When averaged over a sufficiently large number of instances, the
population mean and expectation value are equal.)

If $z$ is an incoherent sum of two statistically
independent and normally-distributed complex circular scalar fields,
i.e.\ $z=z_A + z_B$, then $\mean{\instI} = \mean{\instI_A} +
\mean{\instI_B}$, where $\instI_A=z_A^*z_A$, $\instI_B=z_B^*z_B$, and
\begin{equation}
\varsigma_\instI^2 = \mean{\instI}^2 = \left(\mean{\instI_A} + \mean{\instI_B}\right)^2 
          = \varsigma_A^2 + \varsigma_B^2 + 2\varsigma_A\varsigma_B,
\label{eqn:variance_exponential_intensity}
\end{equation}
where $\varsigma_A=\mean{\instI_A}$ and $\varsigma_B=\mean{\instI_B}$
are the standard deviations of the exponentially distributed
intensities, $\instI_A$ and $\instI_B$ respectively.
%
%
This elementary result demonstrates the fact that, when two
statistically independent fields are superposed, the variance of the
intensity of the resulting sum is not equal to the sum of the
variances of the intensities of the two fields.
The variance of the intensity
$\varsigma_\instI^2=\mean{\instI^2}-\mean{\instI}^2$, where the second
moment $\mean{\instI^2}$ is equal to the fourth-order moment of $z$;
therefore, this example also demonstrates that the fourth-order moment
of a sum of two random variables is not equal to the sum of their
fourth moments.

To derive a general expression for the variance of the instantaneous
intensity of superposed fields that are not necessarily normally
distributed, it is useful to exploit the relationships between the
higher-order moments and cumulants of random variables.
Following \cite{agl96a} and \cite{eok10}, the characteristic function
of a complex-valued circular random variable $z$ is defined as
\begin{eqnarray}
\Phi_z(Z) & \equiv & \mean{ \exp \left( i\re{z^*Z} \right) } \label{eqn:characteristic_function} \\
          & \equiv & \int e^{i\re{z^*Z}} p_z(z) dz
\end{eqnarray}
where $i=\sqrt{-1}$, $p_z(z)$ is the probability density function of $z$, and
$Z$ is the Fourier conjugate of $z$.
That is, the above integral is equivalent to the two-dimensional Fourier
transform of $p_z(z)$ along the real and imaginary components of $z$,
for which the Fourier conjugate variables are the real and imaginary
parts of $Z$, respectively.
The $n^\mathrm{th}$-order moment of $z$
\begin{equation}
\mu_{r;s}(z) \equiv \mean{z^r z^{*s}} 
 = \left. 
    \left( \frac{2}{i} \right)^{r+s} 
    \frac{\partial^{r+s} \Phi_z}{\partial Z^s \partial Z^{*r}}
   \right\vert_{Z=0},
\end{equation}
where $r$ and $s$ are positive integers such that $r+s=n$; the above
equation identifies the moments as the coefficients in a power series
expansion of the characteristic function.
For a given order $n$, there are $n+1$ different moments of a
complex-valued random variable; however, for a circular complex variate,
the only non-zero moments are those for which $r=s$.

If $z=z_A+z_B$ is the sum of two statistically independent circular
complex variates $z_A$ and $z_B$, then the probability density
function of z is the convolution of the probability density functions
of $z_A$ and $z_B$; i.e.
\[
p_z (z) = p_A(z) * p_B(z).
\]
Furthermore, as the Fourier transform of the probability density
function, the characteristic function of $z$ is the product of the
characteristic functions of $z_A$ and $z_B$; i.e.
\[
\Phi_z(Z) = \Phi_A(Z) \Phi_B(Z).
\]
Finally, the natural logarithm of the characteristic
function of $z$, known as the secondary characteristic function,
\begin{equation}
\Psi_z(Z) \equiv \log\left[ \Phi_Z(Z) \right]
\end{equation}
is simply the sum of the secondary characteristic functions of $z_A$
and $z_B$; i.e.
\[
\Psi_z(Z) = \Psi_A(Z) + \Psi_B(Z).
\]
This important property is passed on to the cumulants, which are the
coefficients in a power series expansion of the secondary
characteristic function; i.e.
\begin{equation}
\kappa_{r;s}(z) \equiv
 \left.
   \left( \frac{2}{i} \right)^{r+s}
   \frac{ \partial^{r+s} \Psi_z}{\partial Z^s \partial Z^{*r}}
 \right\vert_{Z=0}.
\end{equation}
Therefore, $\kappa_{r;s}(z)=\kappa_{r;s}(z_A)+\kappa_{r;s}(z_B)$.
Using the relationship between the fourth-order cumulants and
moments of circular complex variables \citep[e.g.][]{men91,eok10},
\[
\kappa_{2;2}(z)=\mu_{2;2}(z) - 2 \left[\mu_{1;1}(z)\right]^2
 = \varsigma_\instI^2 - \mean{\instI}^2,
\]
it is trivial to show that
%
%
\begin{equation}
\varsigma_\instI^2 = \varsigma_A^2 + \varsigma_B^2 + 2 \mean{\instI_A}\mean{\instI_B}.
\label{eqn:fourth_moment_intensity}
\end{equation}

This general expression for the variance of the instantaneous
intensity of a scalar field holds regardless of the distribution of
the circular complex variables $z_A$ and $z_B$ and,
in the special case of normally distributed variates,
it is consistent with \Eqn{variance_exponential_intensity}.
Noting that the variance of a random variable is related to its
autocorrelation function at zero lag, the above equation is
consistent with Equation~16 of \citet{ric75}.\footnote
{There is a typographical error in Equation~16 of \citet{ric75}; the
  third term on the right-hand side of this equation should be
  $\langle r_{x^*}(\tau) \rangle \langle r_N(\tau) \rangle$.}


\subsubsection{Complex-valued Random Vectors}
\label{sec:multivariate_statistics}

To study the statistics of the electric field vector requires
multivariate analysis and the most elegant mathematical descriptions
of random vectors employ tensor algebra \citep[e.g.][]{mcc87,agl96a,smi11}.
This is especially apparent when studying higher-order moments,
as alternative approaches to multivariate analysis typically result
in complicated expressions that involve the Kronecker product, the
vectorization operator (which converts $p\times q$ matrices into
$pq$-dimensional vectors), and commutation (also known as permutation)
matrices; e.g.\ see
Equation (50) of \citet{men91}, 
Equation (2.17) of \citet{st96}, and
Equation (2.1.53) of \citet{kr05}.

Following \cite{agl96a}, the $n^\mathrm{th}$-order moments of the
two-dimensional complex-valued transverse electric field vector $\mbf{e}$ 
are given by $n+1$ rank $n$ tensors defined by
\begin{equation}
\mbf{\mu}_{p;q}(\mbf{e}) 
  \equiv \mean{ \mbf{e}^{\otimes p} \otimes \mbf{e}^{\dagger\otimes q} },
\label{eqn:nth_moment}
\end{equation}
where $p$ and $q$ are positive integers, $p+q=n$,
$\mbf{e}^\dagger$ is the Hermitian transpose of $\mbf{e}$,
$\otimes$ indicates the tensor product, and
$\mbf{e}^{\otimes p}$ is the $p^\mathrm{th}$ tensor power of $\mbf{e}$,
which indicates that $\mbf{e}$ enters into the tensor product $p$
times.
(Defined recursively, $\mbf{e}^{\otimes p+1} = \mbf{e}^{\otimes p} \otimes \mbf{e}$.)
As in the previous section, the $n^\mathrm{th}$-order tensor moments
are related to the characteristic function 
\begin{equation}
\Phi_{\mbfs{e}}(\mbf{E}) \equiv \langle \exp \left( i\re{\mbf{e}^\dagger\mbf{E}} \right) \rangle
\end{equation}
by
\begin{equation}
\mbf{\mu}_{p;q}(\mbf{e}) 
  = \left. 
    \left( \frac{2}{i} \right)^{p+q} 
    \nabla^{\otimes q}_{\mbfs{E}} \otimes \nabla^{\otimes p}_{\mbfs{E^\dagger}} \Phi_{\mbfs{e}}
   \right\vert_{\mbfs{E}=0},
\end{equation}
where $\mbf{e}^\dagger\mbf{E}$ represents an inner product and
$\nabla^{\otimes q}_{\mbfs{E}}$ is the $q^\mathrm{th}$-order gradient
with respect to $\mbf{E}$.
Also as for scalar fields, the secondary characteristic function
\begin{equation}
\Psi_{\mbfs{e}}(\mbf{E}) \equiv \log\left[\Phi_{\mbfs{e}}(\mbf{E})\right]
\end{equation}
is used to define the $n^\mathrm{th}$-order tensor cumulants
\begin{equation}
\mbf{\kappa}_{p;q}(\mbf{e}) 
  = \left. 
    \left( \frac{2}{i} \right)^{p+q} 
    \nabla^{\otimes q}_{\mbfs{E}} \otimes \nabla^{\otimes p}_{\mbfs{E^\dagger}} \Psi_{\mbfs{e}}
   \right\vert_{\mbfs{E}=0}.
\end{equation}

The fourth-order cumulants are related to the fourth- and second-order
moments by \citep{car91}\footnote { Section 5.2 of
  \citet{agl96a} concludes with an incorrect expression for this
  relationship. }
\begin{equation}
\mbf{\kappa}_{2;2}(\mbf{e}) 
= \langle\outerBilinear{\mbf{r}}{\mbf{r}}\rangle
- \outerBilinear{\mbf\rho}{\mbf\rho}
- \spinorBilinear{\mbf\rho}{\mbf\rho},
\label{eqn:fourth_complex_cumulant}
\end{equation}
where $\mbf{r}\equiv\mbf{e} \,\otimes\, \mbf{e}^\dagger$ is the instantaneous
coherency matrix, and
\begin{equation}
\mbf{\rho}\equiv\mbf{\mu}_{1;1}(\mbf{e})=\langle\mbf{r}\rangle
\end{equation}
defines the population mean coherency matrix that is typically used to
describe the polarization of electromagnetic radiation \citep{bw70}.
The $\tilde\otimes$ operator represents a tensor product followed by a
transpose over contravariant tensor indeces; i.e.
\begin{eqnarray}
\left\{\outerBilinear{\bf A}{\bf B}\right\}_{ik}^{jl}
& \equiv &
A_i^j B_k^l  \label{eqn:indexOuterBilinear} \\
\left\{\spinorBilinear{\bf A}{\bf B}\right\}_{ik}^{jl}
& \equiv &
A_i^l B_k^j \label{eqn:indexSpinorBilinear} 
\end{eqnarray}
where {\bf A} and {\bf B} are rank 2 tensors (i.e.\ matrices).
\App{comparison_with_car91} describes some minor differences between
the index notation used in the above definitions and the convention
used by \citet{car91}.

\Eqns{fourth_complex_cumulant}{through}{indexSpinorBilinear} form
the mathematical basis from which the covariances between the
 Stokes parameters are derived in the following sections.
First, the above tensor products are transformed to be expressed in
terms of the Stokes parameters associated with matrices {\bf A} and
{\bf B}.
These transformations are then used to express
\Eqn{fourth_complex_cumulant} in terms of the instantaneous and
population mean Stokes parameters of a single source of radiation.
Finally, in \Sec{superposed}, these tensors are used to derive the
covariances between the Stokes parameters in the case of superposed
modes of emission represented by {\bf A} and {\bf B}.



\section{Covariances between the Instantaneous Stokes parameters}
\label{sec:Stokes_statistics}

Although more elegant than its alternative representations, the
complex-valued rank 4 tensor of cumulants defined in
\Eqn{fourth_complex_cumulant} is not immediately amenable to
interpretation.
It remains to transform the equations that describe fourth-order
tensor products of the electric field vector into the equivalent
objects that describe second-order tensor products of the Stokes
parameters.
This is achieved in the following sections by first identifying the
isomorphism between rank 4 tensors in the two-dimensional vector space
of the electric field \rankfour\ and rank 2 tensors in the
four-dimensional vector space of the Stokes parameters \ranktwo.

\newpage
\subsection{Isomorphism between \rankfour\ and \ranktwo}
  
The required mathematical mapping between fourth-order products of
$\mbf{e}$ and second-order products of Stokes parameters derives
from the isomorphism
%
%
between the $2\times2$ vector space of the coherency matrix and the
four-dimensional vector space of the Stokes parameters, as expressed
by the following pair of equations.
\begin{eqnarray}
{\mbf\rho} & = & S_\irow\,\pauli{\irow} / 2    \label{eqn:combination} \\
S_\irow & = & \dc{\pauli{\irow}}{\mbf\rho}     \label{eqn:projection}
\end{eqnarray}
Here, $S_\irow$ are the four Stokes parameters, Einstein notation is used
to imply a sum over repeated indeces, $0\le\irow\le3$, $\pauli{0}$ is
the $2\times2$ identity matrix, $\pauli{1-3}$ are the Pauli matrices,
and the $\mbf{:}$ operator represents tensor double contraction, a
tensor product followed by contraction over two pairs of indeces.
The double contraction of two matrices {\bf A} and {\bf B} yields
a scalar quantity defined by
\begin{equation}
\dc{\bf A}{\bf B} \equiv A_i^j B_j^i = \tr{\bf AB},
\label{eqn:double_contraction_matrices}
\end{equation}
where $\trace$ is the matrix trace operator.
\Eqn{combination}
expresses the coherency matrix as a linear combination of Hermitian
basis matrices; \Eqn{projection} represents the Stokes parameters as
the projections of the coherency matrix onto the basis matrices.  When
$\mbf\rho$ is Hermitian, the Stokes parameters are real-valued.

Using \Eqns{combination}{and}{projection}, any linear transformation
of the coherency matrix $\mbf\rho^\prime=L(\mbf{\rho})$, can be
expressed as an equivalent linear transformation of the associated
Stokes parameters by the Mueller matrix {\bf M}, as defined by
\begin{equation}
 S_\irow^\prime = M_\irow^\icol S_\icol
 = \frac{1}{2} \dc{\pauli{\irow}}{L(\pauli{\icol}) } S_\icol.
\label{eqn:linear_to_Mueller}
\end{equation}
If, for any positive-definite Hermitian matrix $\mbf\rho$, the result
of the linear transformation $L(\mbf\rho)$ is also
positive-definite and Hermitian, then $L$ is called {\it positive}
and its associated Mueller matrix is real-valued.  In general,
the Mueller matrix may be complex-valued.

Let $L_{\bf U}(\mbf{\rho}) = \dc{\bf U}{\mbf\rho}$ represent the
linear transformation of $\mbf\rho$ by a rank 4 tensor {\bf U} with 2
covariant and 2 contravariant indeces, such that the double contraction 
yields a matrix with components given by
\begin{equation}
\left\{ \dc{\bf U}{\mbf\rho} \right\}_i^j
  \equiv U_{ik}^{jl} \rho_l^k.
\label{eqn:double_contraction}
\end{equation}
Substitution of $L_{\bf U}$ into \Eqn{linear_to_Mueller}, followed by
elimination of $S_\icol$, associates with {\bf U} a 
$4\times4$ Mueller matrix,
\begin{equation}
M_\irow^\icol 
= \frac{1}{2} \dc{\pauli{\irow}}{\dc{\bf U}{\pauli{\icol}}}.
\label{eqn:tensor_to_Mueller}
\end{equation}
Likewise, for any $4\times4$ Mueller matrix, there is an associated
rank 4 tensor,
\begin{equation}
{\bf U} = \frac{1}{2} M_\irow^\icol \outerBilinear{\pauli{\irow}}{\pauli{\icol}}.
\label{eqn:Mueller_to_tensor}
\end{equation}
\Eqn{tensor_to_Mueller} expresses the components of a Mueller matrix
as the double projections of a rank 4 tensor {\bf U} onto the Hermitian
basis matrices.
\Eqn{Mueller_to_tensor} represents {\bf U} as a linear combination of
the 16 basis tensors formed by all possible tensor products of the 4
Hermitian basis matrices.
Combined, these equations establish the required isomorphism between
\rankfour\ and \ranktwo.
A similar mapping between a Mueller matrix and its associated target
coherency matrix was derived by \cite{clo86} using Kronecker
products of the Hermitian basis matrices and the matrix trace
operator.


\subsection{Tensor products of the Stokes parameters}

Although \Eqns{tensor_to_Mueller}{and}{Mueller_to_tensor} are derived
by considering equivalent linear transformations of the coherency
matrix and Stokes four-vector, they can be used to map any object from
\rankfour\ to \ranktwo\ (and vice versa).
Bearing in mind the objective to convert \Eqn{fourth_complex_cumulant}
into an equivalent form that is expressed in terms of the Stokes parameters,
\Eqn{tensor_to_Mueller} is first used to convert the tensor
products $\outerBilinear{\bf A}{\bf B}\in\rankfour$ and
$\spinorBilinear{\bf A}{\bf B}\in\rankfour$ defined by
\Eqns{indexOuterBilinear}{and}{indexSpinorBilinear} into the
equivalent tensor products
$\outerBilinear{A}{B}\in\ranktwo$ and
$\spinorBilinear{A}{B}\in\ranktwo$, respectively, where
$A$ and $B$ are the Stokes parameters associated
with {\bf A} and {\bf B}.
Setting either ${\bf U}=\outerBilinear{\bf A}{\bf B}$ or
${\bf U}=\spinorBilinear{\bf A}{\bf B}$ in \Eqn{tensor_to_Mueller},
the following
transformation properties \citep{car91}
\begin{eqnarray}
\left( \outerBilinear{\bf A}{\bf B} \right) \mbf{:}\,{\bf C}
& = &
{\bf A}\left(\dc{\bf C}{\bf B}\right) \label{eqn:outerBilinear} \\
\left( \spinorBilinear{\bf A}{\bf B} \right) \mbf{:}\,{\bf C}
& = &
{\bf A}{\bf C}{\bf B} \label{eqn:spinorBilinear} 
\end{eqnarray}
are applied to
yield
\begin{eqnarray}
\left\{ \outerMueller{\bf A}{\bf B} \right\}_\irow^\icol
& = &  
\frac{1}{2}\left(\dc{\pauli{\irow}}{\bf A}\right)\left(\dc{\pauli{\icol}}{\bf B}\right)
\label{eqn:outerMueller} \\
\left\{ \spinorMueller{\bf A}{\bf B} \right\}_\irow^\icol
& = &
\frac{1}{2}\dc{\pauli{\irow}}{\left({\bf A}\,\pauli{\icol}\,{\bf B}\right)}
 \label{eqn:spinorMueller}
\end{eqnarray}
Using \Eqn{projection}, \Eqn{outerMueller} is related to the 
tensor product of the Stokes parameters; i.e.
\begin{equation}
A \otimes B = 2 \outerMueller{A}{B}
\label{eqn:outerStokes}
\end{equation}
where 
\begin{equation}
\left\{ A \otimes B \right\}_\irow^\icol \equiv A_\irow B_\icol.
\end{equation}
Similarly, as shown in \App{linear_to_Mueller}, \Eqn{spinorMueller}
yields
\begin{equation}
  \spinorBilinear{A}{B}=
  \frac{1}{2}\left(A \otimes B + B \otimes A
  - {\mbf\eta}\Linner{A}{B} + i\,\exterior{A}{B}\right),
\label{eqn:spinorStokes}
\end{equation}
where $\mbf\eta$ is the Minkowski metric tensor with signature
$(+,-,-,-)$,
\begin{equation}
\Linner{A}{B} \equiv \eta^{\irow\icol} A_\irow B_\icol = A_0B_0 - \mbf{A \cdot B}
\end{equation}
is the covariant inner product of $A$ and $B$, and
$\exterior{A}{B}$ is the anti-symmetric covariant exterior product
of the Stokes parameters, defined in \Eqn{exterior}.

\Eqns{outerStokes}{and}{spinorStokes} are the required tensor products
of the Stokes parameters that yield $4\times4$ matrix representations
of the tensor products defined by
\Eqns{indexOuterBilinear}{and}{indexSpinorBilinear}, respectively.


\subsection{The Stokes cumulant}

\begin{figure}
\centerline{\includegraphics[width=50mm]{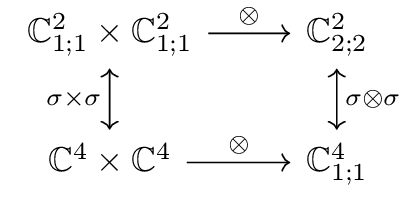}}
\caption{\label{fig:commutative}
Commutative diagram representing the linear and bilinear mappings
between the vector spaces used in this work.  In this diagram,
$\mathbb{C}^2_{1;1}$ represents the vector space of two-dimensional
rank 2 tensors such as the coherency matrix and $\mathbb{C}^4$
represents the vector space of four-dimensional vectors such as the
Stokes parameters.
The linear map $\sigma$ represents the isomorphism between these
vector spaces as embodied in \Eqns{combination}{and}{projection}.
In the top row, $\mathbb{C}^2_{1;1}\times\mathbb{C}^2_{1;1}$
represents a Cartesian product, the vector space of ordered pairs
$\left({\bf A}, {\bf B}\right)$ where ${\bf A}\in\mathbb{C}^2_{1;1}$
and ${\bf B}\in\mathbb{C}^2_{1;1}$; the tensor product defined by
\Eqn{indexOuterBilinear} maps these ordered pairs into rank four
tensors ${\bf A}\otimes{\bf B}\in\rankfour$.
Similarly, in the bottom row, $\mathbb{C}^4\times\mathbb{C}^4$
represents the vector space of ordered pairs $\left(A, B\right)$ where
$A\in\mathbb{C}^4$ and $B\in\mathbb{C}^4$; the tensor product defined
by \Eqn{outerStokes} maps these ordered pairs into rank 2 tensors
$A\otimes B\in\ranktwo$.
On the left edge, $\sigma\times\sigma$ represents the Cartesian
product of the $\sigma$ operator with itself.  When
$\sigma\times\sigma$ is applied to an ordered pair, it produces a new
ordered pair in which the $\sigma$ operator has been applied
separately to each of the two elements; i.e.
%
$\left(\sigma\times\sigma\right)\left(A,B\right)
   = \left(\sigma A,\sigma B\right)$.
%
On the right edge, $\sigma\otimes\sigma$ represents the isomorphism
between \ranktwo\ and \rankfour, as embodied in
\Eqns{tensor_to_Mueller}{and}{Mueller_to_tensor}.
When $\sigma\otimes\sigma$ is applied to a tensor product it produces
a new tensor product in which the $\sigma$ operator has been applied
separately to each of the two operands; i.e.
%
$\left(\sigma\otimes\sigma\right)\left(A\otimes B\right)
   = \left(\sigma A\right)\otimes\left(\sigma B\right)$.
%
Owing to the universal property of the tensor product, the
$\sigma\otimes\sigma$ operator is unique.  A similar commutative
diagram can be drawn in which the $\otimes$ operators on the top and
bottom edges are replaced by the $\stimes$ operator.}
\end{figure}

The results of the previous two sub-sections are summarized in the
commutative diagram shown in \Fig{commutative}.
Although the mappings shown in this diagram have been derived and
discussed in the context of the Stokes parameters and coherency
matrix, they are purely algebraic transformations between vector
spaces and are completely independent of any consideration of
statistical moments.
In this section, these mappings are applied to convert
\Eqn{fourth_complex_cumulant} into an equivalent relation between the
cumulants and moments of the Stokes parameters.
Setting ${\bf U} = \mbf{\kappa}_{2;2}(\mbf{e})$ in
\Eqn{tensor_to_Mueller} and converting each of the three terms on the
right hand side of \Eqn{fourth_complex_cumulant} using
\Eqns{outerStokes}{and}{spinorStokes} yields the Stokes cumulant matrix
\begin{equation}
  {\bf Q} = \langle\outerBilinear{s}{s}\rangle
  - \outerBilinear{S}{S} - \spinorBilinear{S}{S},
\label{eqn:Stokes_cumulant}
\end{equation}
where lower-case $s$ represents the instantaneous Stokes
four-vector with components defined by
\[
s_\irow\equiv\dc{\mbf{r}}{\pauli{\irow}}=\mbf{e}^\dagger\pauli{\irow}\mbf{e},
\]
and upper-case $S\equiv\mean{s}$ represents the population
mean Stokes parameters.  Noting that $S\wedge S=0$,
\begin{equation}
\spinorBilinear{S}{S}
  = S \otimes S - \frac{1}{2} \mbf{\eta}\inv{S},
\label{eqn:Lorentz_transformation}
\end{equation}
where $\inv{S}$ is the invariant interval of the Stokes
four-vector \citep[e.g.][]{bar63,scr+84,bri00} defined by
\begin{equation}
\inv{S} \equiv \Linner{S}{S} = S_0^2 - |\mbf{S}|^2.
\end{equation}
Here, the Stokes four-vector is separated into the total intensity
$S_0$ and the polarization vector, $\mbf{S}=(S_1,S_2,S_3)$.

The real-valued $4\times4$ matrix {\bf Q} represents the fourth-order
cumulants of the electric field vector as second-order moments of the
instantaneous Stokes parameters; it is used in \Sec{superposed} to
determine the covariances between the Stokes parameters when two
sources of radiation are superposed.
Even when considering only a single source of radiation, the Stokes
cumulant matrix can be used to compute the covariances between the
instantaneous Stokes parameters, which are defined by the $4\times4$
covariance matrix,
\begin{equation}
  {\bf C} \equiv \langle\outerBilinear{s}{s}\rangle - \outerBilinear{S}{S}.
  \label{eqn:covariance-instantaneous}
\end{equation}
Substitution of the above into \Eqn{Stokes_cumulant} yields
\begin{equation}
  {\bf Q} = {\bf C} - \spinorBilinear{S}{S}.
\end{equation}
If the components of the electric field are jointly drawn from a
circular complex-valued multivariate normal distribution
\citep{goo63}, then all cumulants above second order are equal to
zero and ${\bf C}=\spinorBilinear{S}{S}$, which is
consistent with the results of \cite{bb92} and \cite{van09}.
In this special case, the covariances between the instantaneous Stokes
parameters are completely defined by the population mean Stokes
parameters.
\citet{van09} argued that ${\bf C}=\spinorBilinear{S}{S}$ regardless of the
distribution of the electric field; however, as shown by
\Eqn{Stokes_cumulant} and discussed in more detail in
\Sec{modulation}, this assertion is incorrect and the relation holds
only in the special case that the electric field vector is complex
circular normal.


\section{Covariances between the Sample Mean Stokes Parameters}
\label{sec:modes}

A statistical description of the sample mean Stokes parameters begins
with the definition of a {\em Stokes sample}: a finite sample of $n$
instances of the electric field vector from which the sample mean
coherency matrix $\bar{\mbf\rho}$ and sample mean Stokes parameters
$\bar{S}$ are computed via
\begin{equation}
\label{eqn:local_rho}
\bar{\mbf\rho} \equiv \frac{1}{n} \sum_{i=1}^n \mbf{e}_i \otimes \mbf{e}_i^\dagger
                = \frac{1}{2} \bar{S}_\irow\pauli{\irow}.
\end{equation}
Over a population of independent estimates of $\bar S$, each drawn
from mutually exclusive Stokes samples,\footnote{
For example, the population could consist of the sample mean Stokes
parameters for a given phase bin drawn from a series of
sub-integrations of the average pulse profile.}
the covariances between the sample mean Stokes parameters are defined
by the $4\times4$ covariance matrix,
\begin{equation}
{\bar{\bf C}} = \langle \bar{S} \otimes \bar{S} \rangle - S \otimes S,
  \label{eqn:covariance-sample}
\end{equation}
where $S\equiv\mean{\bar{S}}$ represents the population mean Stokes
parameters.\footnote {The population mean of the sample
  mean Stokes parameters is equal to the population mean of the
  instantaneous Stokes parameters.}
If the instances of the electric field vector are statistically
independent and identically distributed, then ${\bar{\bf C}} = {\bf C}n^{-1}$,
where {\bf C} is the covariance matrix of the instantaneous Stokes
parameters defined in \Eqn{covariance-instantaneous} and $n$ is the
number of independent instances of the electric field in each Stokes
sample.

\begin{figure}
\centerline{\includegraphics[angle=-90,width=60mm]{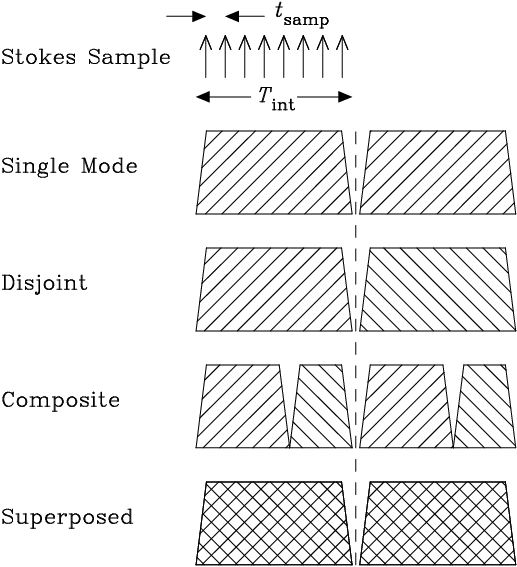}}
\caption{\label{fig:regimes} Depiction of different Stokes sample
  types in the time domain.  A Stokes sample (represented as a series
  of $n=8$ delta functions) consists of a sequence of instances of the
  electric field sampled at regular intervals separated by
  $t_\mathrm{samp}$ and spanning an integration interval defined by
  $T_\mathrm{int}$.  Two independent Stokes samples are shown
  separated by a dashed vertical line.  Emission in one mode is
  represented by a trapezoid filled with hatch lines of a single
  orientation; emission in the other mode is represented using hatch
  lines rotated by 90 degrees.  Each disjoint Stokes sample is
  entirely comprised of instances drawn exclusively from only one
  population or the other.  Composites Stokes samples include
  instances drawn from both populations.  When the modes are
  superposed, a cross hatch pattern represents the new polarization
  state that arises when electric field instances drawn from both
  populations are added together before the instantaneous Stokes
  parameters are computed.
  Although two Stokes samples are shown side-by-side, they need not
  necessarily be contiguous in the time domain.
  For example, in single-pulse studies, each Stokes sample might be
  integrated over a fixed range of pulsar longitude such that the
  start time of each Stokes sample is temporally separated from the
  next by the pulsar spin period.
  In this case, the sample mean Stokes parameters record the
  polarization state of the pulsar emission over that longitude range
  as a function of pulse number (integer turns of the pulsar).
}
\end{figure}

\Fig{regimes} illustrates the special case of Stokes samples drawn in
the time domain; however, it should be noted that the analysis
presented in this paper applies in any domain (temporal, spectral, or
spatial).
In addition to emission from a single source (or mode) of radiation,
this figure depicts the combination of electromagnetic waves from two
sources (or two modes of emission from a single source) in the three
special cases outlined in the introduction (disjoint, superposed, and
composite samples).
These classifications are highly idealized and in reality the
distinction between statistical regimes may not be so clean.
For example, the scale for switching between mutually exclusive modes
could be variable and span intervals that are both smaller and
greater than the integration interval of the instrument; in this case,
the source would be described by a mixture of disjoint and
composite samples.
Similarly, mutually exclusive modes can become partially superposed by
processes that introduce temporal, spatial, or spectral coherence,
such as passage through a linear time-invariant system (such as the
interstellar medium) that is characterized by an impulse response
function with a duration that exceeds the timescale for switching
between modes.
Nevertheless, these idealizations provide a sound conceptual framework
on which to base more realistic models.

In the following sections, $\mbf{e}_A$ and $\mbf{e}_B$ are the
electric field vectors associated with two modes of emission; the
modes are strictly stationary and have population mean Stokes
parameters, $A$ and $B$.
The two modes may have completely different statistical distributions;
e.g. the intensity of mode A could be log-normally distributed and that
of mode B could be exponentially distributed.
It is assumed that instances of the electric field from a single mode
are statistically independent and identically distributed; therefore
$\bar{\bf C}_A={\bf C}_A n^{-1}$ 
and 
$\bar{\bf C}_B={\bf C}_B n^{-1}$
describe the covariances between the sample mean Stokes parameters
after averaging over a Stokes sample of $n$ instances of the electric
field drawn exclusively from either mode A or mode B, respectively.
\Secs{disjoint}{through}{composite} present general expressions for
the covariances between the Stokes parameters in the three statistical
regimes, illustrative examples are presented in \Sec{example}, and in
\App{comparison_with_crb78} it is demonstrated that composite samples
of unresolved disjoint modes may have been incorrectly identified as
superposed modes in previous studies.


\subsection{Disjoint Samples}
\label{sec:disjoint}

If the modes are mutually exclusive and the interval over which the
Stokes parameters are integrated is smaller than the scale over which
mode switching occurs, then the modes are resolved.
In this case, every instance of the electric field vector in a given
Stokes sample is drawn exclusively from only one of the two
populations and the sample mean Stokes parameters are given by either
$\bar A$ or $\bar B$.
When averaged over many Stokes samples, the population mean
Stokes parameters are given by the weighted mean of the population
mean Stokes parameters of each mode,
$S = F A + (1-F) B$,
where $F$ is the fraction of Stokes samples that occur in mode A.
Although $F$ might reasonably fluctuate (e.g. as a function of
time), only the ideal case in which $F$ remains constant is
considered for now.
If the modes are statistically independent, then the covariance matrix
of the sample mean Stokes parameters is given by
%
%
%
%
%
%
%
%
%
%

\begin{equation}
\bar{\bf C}_d = F \bar{\bf C}_A + (1-F) \bar{\bf C}_B + F(1-F) \outerDiff{A}{B},
\label{eqn:covariance-disjoint}
\end{equation}
where $\outerDiff{A}{B} = (A-B)\otimes(A-B)$ is the second tensor
power of the differences between the population mean Stokes parameters
of the modes.
As shown in \App{comparison_with_crb78},
the above equation is consistent with the definition of disjoint modes
presented in equation (5) of \cite{crb78}.
%


\subsection{Superposed Samples}
\label{sec:superposed}

When the electromagnetic wave modes are superposed, each instance of
the electric field vector in a Stokes sample is given by the sum,
$\mbf{e}=\mbf{e}_A+\mbf{e}_B$.
For superposed samples, the population mean coherency matrix is
given by
\begin{equation}
\mbf{\rho} = \mbf{\rho}_A + \mbf{\rho}_B + \mbf{\chi} + \mbf{\chi}^\dagger
\label{eqn:coherency-superposed}
\end{equation}
where $\mbf{\rho}_A$ and $\mbf{\rho}_B$ are the coherency matrices
of $\mbf{e}_A$ and $\mbf{e}_B$, and
\begin{equation}
\mbf{\chi} \equiv \mean{\mbf{e}_A\,\mbf{e}_B^\dagger}
\end{equation}
is the cross-coherency matrix that describes the coherence of the
polarized modes. 
In the special case of an incoherent sum, the population mean
$\mbf\chi = 0$; however, in each finite Stokes sample, the sample mean
$\bar{\mbf\chi}$ is not exactly zero and the variance of the sample
mean cross-coherency contributes to the fourth-order moments of the
electric field.

If the modes are statistically independent, then the population mean
Stokes parameters are $S=A+B$ and the Stokes cumulant of the sum is
${\bf Q}={\bf Q}_A+{\bf Q}_B$.
Using \Eqn{Stokes_cumulant}, the covariances between the
sample mean Stokes parameters of the sum are
\begin{equation}
\bar{\bf C}_s = \bar{\bf C}_A + \bar{\bf C}_B + n^{-1} \outerSymm{A}{B},
\label{eqn:covariance-superposed}
\end{equation}
where
\begin{eqnarray}
\outerSymm{A}{B} &=& \spinorBilinear{A}{B}+\spinorBilinear{B}{A} \nonumber \\
& = & A \otimes B + B \otimes A - {\mbf\eta}\Linner{A}{B}
\label{eqn:symmetric}
\end{eqnarray}
is twice the symmetric part of \spinorBilinear{A}{B}; it describes
the cross-covariance between the instantaneous Stokes parameters of
the modes and depends on only the population mean Stokes parameters of
the modes.
As shown in \App{comparison_with_crb78}, \Eqn{covariance-superposed} is
consistent with the definition of superposed modes presented in
equation (5) of \cite{crb78} only in the special case of 100\%
polarized modes and only when considering the instantaneous Stokes
parameters.

The cross-covariance $\outerSymm{A}{B}$ is a function of only the population
mean Stokes parameters and does not explicitly depend on unique
statistical degrees of freedom for each of the modes, as asserted in
the erratum of \cite{van09}.
As an aside, note that the variance of the instantaneous ($n=1$) total
intensity predicted by the above equation is consistent with the variance
of the instantaneous intensity of superposed scalar fields.
This can be shown by transforming the electric field vector by a
singular Jones matrix that reduces its dimension to that of a single
scalar field.
In this basis, all sources are observed to be 100\% polarized, such
that $A \cdot B = 0$ and $\left\{\bar{\bf C}_s\right\}_0^0$ of
\Eqn{covariance-superposed} reduces to $\varsigma_\instI^2$ of
\Eqn{fourth_moment_intensity}.


\subsection{Composite Samples}
\label{sec:composite}

When only one mode contributes at any instant and both modes
contribute to the signal in the interval over which the sample mean
Stokes parameters are integrated, then the disjoint modes
are unresolved.
In this case, each Stokes sample is a union of sub-samples drawn from
two mutually exclusive populations
and the composite sample mean Stokes parameters are equal to the
weighted average of the sample mean Stokes parameters of each mode;
i.e.\ $\bar{S} = f \bar{A} + (1-f) \bar{B}$,
where $f$ is the constant fraction of samples that occur in mode A
in each Stokes sample.
It is also reasonable that $f$ might fluctuate between Stokes samples;
here, only the ideal case in which $f$ remains constant is considered.
If the modes are statistically independent, then the covariance matrix
of the sample mean Stokes parameters is given by
%
%
%
%
%
%
\begin{equation}
\bar{\bf C}_c = f \bar{\bf C}_A + (1-f) \bar{\bf C}_B.
\label{eqn:covariance-composite}
\end{equation}
As shown in \App{comparison_with_crb78},
the above equation is consistent with the definition of mode
superposition presented in equation (5) of \cite{crb78}.
It is also consistent with the definition of superposed modes
described in Section 2.1 of \cite{ms98}
%
and with the definition of an incoherent sum described in Section 4.3
of \cite{van09}.  In the erratum to the latter work, \cite{van10}
redefines an incoherent sum that is consistent with classical wave
superposition, as defined in the previous section; however, the
distinction between superposed and composite samples was not
recognized in this previous work.

\onecolumngrid
\begin{deluxetable*}{lccc}
\tablecaption{Regimes of Mode Combination\label{tab:regimes}}
\tablehead{\colhead{ Property } & \colhead{ Superposed } & \colhead{ Composite } & \colhead{ Disjoint } } 
\startdata
Electric Field Instances
 & $\mbf{e}\in\{\mbf{e}_A+\mbf{e}_B \}_n$
 & $\mbf{e}\in\{\mbf{e}_A\}_{fn}\cup\{\mbf{e}_B\}_{(1-f)n}$
 & $\mbf{e}\in\{\mbf{e}_A\}_n\,\veebar\,\mbf{e}\in\{\mbf{e}_B\}_n$ \\ [2mm]
Sample Mean Stokes Parameters
 & $\bar{S}\in\{\bar{A}+\bar{B}\}_N$
 & $\bar{S}\in\{f\bar{A}+(1-f)\bar{B}\}_N$
 & $\bar{S}\in\{\bar{A}\}_{FN}\cup\{\bar{B}\}_{(1-F)N}$ \\ [2mm]
Covariances of Sample Means
 & $\bar{\bf C}_A + \bar{\bf C}_B + n^{-1}\outerSymm{A}{B}$
 & $f\bar{\bf C}_A + (1-f)\bar{\bf C}_B$
 & $F\bar{\bf C}_A + (1-F)\bar{\bf C}_B + F(1-F)\outerDiff{A}{B}$ \\ [2mm]
Incoherent Sum & pre-detection & post-detection & N/A \\ [2mm]
Mutual Exclusivity & N/A & unresolved & resolved
\enddata
\end{deluxetable*}


\subsection{Illustrative Examples}
\label{sec:example}


The three regimes of mode combination presented in the previous
sub-sections are summarized in \Tab{regimes}, where
$\{\mbf{e}\}_n\equiv\{\mbf{e}_1,\mbf{e}_2,\ldots\mbf{e}_n\}$ is a
Stokes sample of $n$ instances of the electric field vector,
$\{\bar{S}\}_N$ is a population of $N\to\infty$ instances of sample
mean Stokes parameters, $\cup$ is the set union operator, and
$\veebar$ is the exclusive disjunction.
The last two rows of this table highlight some of the semantic and
conceptual similarities between the three statistical regimes.
The term ``incoherent sum'' is used to describe both the
pre-detection classical wave superposition of statistically
independent signals (e.g.\ astronomical source plus receiver noise)
and the post-detection integration of flux densities
(e.g.\ integration over time and radio frequency).
A source that switches between mutually exclusive states will result
in either composite or disjoint samples, depending on the interval
over which the sample mean is integrated and the characteristic scale
for mode switching.

\begin{figure}
\centerline{\includegraphics[angle=-90,width=60mm]{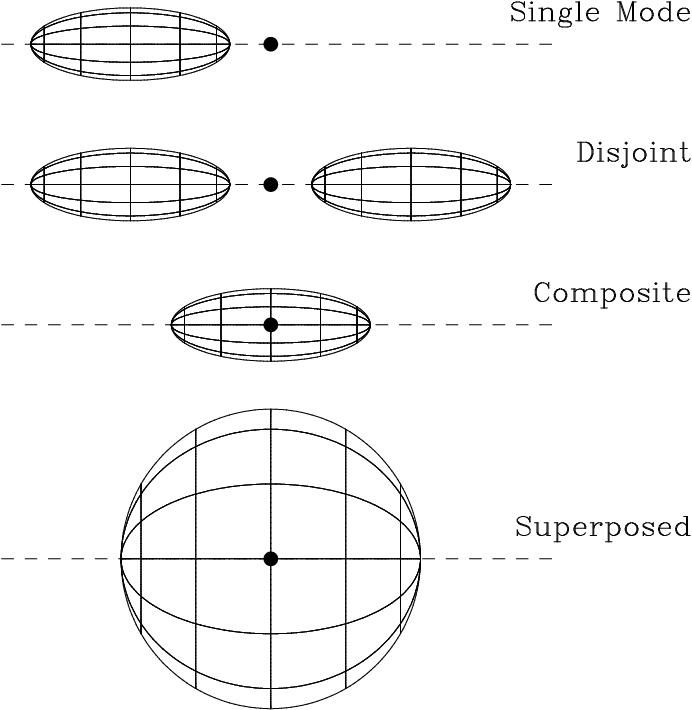}}
\caption{\label{fig:modes} Distributions of the sample mean Stokes
  polarization vector in each regime of orthogonal mode combination.
  In each row, the origin (unpolarized flux) is marked by a dot and
  the principal axis (defined by the population mean polarization
  vectors of the modes) is marked by the dashed line that runs
  horizontally through the origin.  All of the spheroids are symmetric
  under rotation about the principal axis, the axes of the spheroids
  are proportional to the standard deviations of the components of the
  sample mean Stokes polarization vector
  $\bar{\mbf{S}}=(\bar{S}_1,\bar{S}_2,\bar{S}_3)$, and the axial
  ratios are defined by the degree of polarization of the population
  mean Stokes parameters, $p=|\mbf{S}|/S_0$.}
\end{figure}

To illustrate the fundamental differences between the three types of
Stokes sample, consider the incoherent sum of orthogonally polarized
modes with normally distributed electric fields, equal population mean
intensities $I$ and equal degrees of polarization $p$.
Note that orthogonally polarized states have anti-parallel population
mean Stokes polarization vectors and, without any loss of generality,
assume that the basis in which the Stokes parameters are measured is
the natural basis defined by the modes, such that $S_2 = S_3 = 0$.
Furthermore, for both composite and disjoint samples, the modes occur
with equal frequency (i.e. $f=F=0.5$) such that, for all three
sample types, the resulting signal is completely depolarized;
i.e.\ the
population mean Stokes polarization vector $\mbf{S}=0$.
Finally, assume that each Stokes sample is sufficiently large that
the distributions of the sample mean Stokes parameters are well
approximated by a multivariate normal distribution.
(This is a reasonable assumption in single-pulse studies because each
phase bin typically spans thousands of instances of the electric
field.)
Note that the population mean Stokes parameters provide no
information about the manner in which the signals have been combined.
However, as depicted in \Fig{modes}, the three-dimensional
distributions of the sample mean Stokes polarization vector are
fundamentally different in each case.

These differences are also evident in the structure of the $4\times4$
matrix of covariances between the sample mean Stokes parameters.
When only a single source contributes, ${\bf C}=\spinorBilinear{S}{S}$
and in the natural basis \Eqn{Lorentz_transformation} yields
\begin{equation}
\label{eqn:natural_covariance}
\bar{\bf{C}}_\mathrm{single} = \frac{1}{2n} \left(\begin{array}{cccc}
\norm{S}^2 &  2 I^2p    &  0 & 0 \\
2 I^2p     & \norm{S}^2 &  0 & 0 \\
0 & 0 & S^2 & 0 \\
0 & 0 & 0 & S^2 \\
\end{array}\right),
\end{equation}
where the Euclidean norm
\begin{equation}
\norm{S}^2 \equiv S_0^2 + |\mbf{S}|^2.
\end{equation}
In this example, $\norm{S}^2=I^2(1+p^2)$ and $S^2=I^2(1-p^2)$.  Note that the
principal polarization $S_1$ is covariant with the total intensity
$S_0$ and that the variances of $S_0$ and $S_1$ are equal.
Furthermore, the variances of $S_2$ and $S_3$ are equal and less than
the variances of $S_0$ and $S_1$; therefore, the distribution of the
sample mean Stokes polarization vector is described by a prolate
spheroid that is rotationally symmetric about the population mean
Stokes polarization vector.
The axial ratio of this spheroid, $\epsilon=((1+p^2)/(1-p^2))^{1/2}$,
is completely defined by the degree of polarization of the population
mean Stokes parameters.

For disjoint samples, the bimodal distribution of sample mean Stokes
parameters is no longer accurately described by a multivariate
normal distribution.
The difference between the population mean polarization
vector of each mode (i.e.\ the distance between the centres of each
prolate spheroid, $2Ip$) adds to the estimated variance along the
principal axis.
This is readily seen by expressing \Eqn{covariance-disjoint} in the
natural basis, which yields
\[
\bar{\bf{C}}_\mathrm{disjoint} = \frac{1}{2n} \left(\begin{array}{cccc}
\norm{S}^2       & 0 & 0 & 0 \\
0 & \norm{S}^2  + 2 n I^2 p^2     &  0 & 0 \\
0 & 0 & S^2 & 0 \\
0 & 0 & 0 & S^2 \\
\end{array}\right).
\]
Whereas the sizes of the disjoint prolate spheroids are inversely
proportional to the square root of the sample size, the distance
between their centres is not decreased by integration.

For superposed samples, the instantaneous intensity is doubled and the
cross-covariance between the modes yields a hyperspherically symmetric
distribution described by \Eqn{covariance-superposed},
\[
\bar{\bf{C}}_{\mathrm{superposed}} = \frac{1}{2n} \left(\begin{array}{cccc}
4I^2 & 0 & 0 & 0 \\
0 & 4I^2 &  0 & 0 \\
0 & 0 & 4I^2 & 0 \\
0 & 0 & 0 & 4I^2 \\
\end{array}\right).
\]
In this example, superposition yields unpolarized radiation, for which
the distribution of sample mean Stokes parameters is expected to be
hyperspherically symmetric when the voltage is normally distributed.

For composite samples, integration over an equal number of instances
of each mode yields a prolate spheroidal distribution of the sample
mean Stokes polarization vector with the same dimensions as those of
the original modes, now centred on the origin; i.e., from
\Eqn{covariance-composite},
\[
\bar{\bf{C}}_{\mathrm{composite}} = \frac{1}{2n} \left(\begin{array}{cccc}
\norm{S}^2 & 0 & 0 & 0 \\
0 & \norm{S}^2 &  0 & 0 \\
0 & 0 & S^2 & 0 \\
0 & 0 & 0 & S^2 \\
\end{array}\right).
\]
In this example, when orthogonal modes are combined, the principal
polarization is no longer covariant with the total intensity
(i.e. $C_0^1=C_1^0=0$) because the two modes contribute equally and
oppositely to this term.
For both disjoint and composite samples, the axial ratio
$\epsilon=\varsigma_1/\varsigma_2$ that is inferred from the
covariance matrix is greater than unity.
As this is inconsistent with the degree of polarization of the
population mean Stokes parameters, it potentially serves as an
experimental indicator of mutually exclusive modes. \\ [5mm]

%


\section{Interpretation}
\label{sec:interpretation}

Given only the population mean Stokes parameters, the three regimes of
mode combination (disjoint, superposed, and composite samples)
described in the previous section cannot be distinguished;
however, in principle, they may be differentiated via the covariances
between the Stokes parameters.
A $4\times4$ symmetric and real-valued covariance matrix contains 10
unique elements; using principal component analysis, these may be
reduced to 7 numbers of interest as follows.
First, conformably partition the $4\times4$ covariance matrix into 
the variance of the total intensity,
the $3\times3$ covariance matrix of the Stokes polarization vector, and
the 3-dimensional vector of covariances between the total intensity and the
components of the polarization vector.
%
Then, project the Stokes parameters onto the basis defined by the
eigen decomposition of the $3\times3$ covariance matrix of the Stokes
polarization vector.
This diagonalizes the $3\times3$ partition, leaving only four
variances $\varsigma_\irow^2=C_\irow^\irow$ along the diagonal and three covariances
$C_0^\srow\,(=C_\srow^0)$ in the first row (and column).
In this eigenbasis, assuming normally distributed electric field
components, the following conditions hold.

1. Both a single mode and superposed modes produce prolate spheroidal
distributions of the sample mean Stokes polarization vector such that
the primary axis of the distribution (as defined by the eigenvector
with the largest eigenvalue) is aligned with the population mean
Stokes polarization vector, the distribution is symmetric under
rotation about this axis ($\varsigma_2=\varsigma_3$), and the axial
ratio
$\epsilon\equiv\varsigma_1/\varsigma_{2}=((1+p^2)/(1-p^2))^{1/2}$,
where $p=|\mbf{S}|/S_0$ is the degree of polarization of the
population mean Stokes parameters.
The standard deviation of the primary polarization is
equal to that of the total intensity (i.e.\ $\varsigma_1=\varsigma_0$) and
the total intensity is uncorrelated with the minor polarizations
(i.e.\ $C_0^2=C_0^3=0$).

2. Mutually exclusive orthogonally polarized modes (both disjoint and
composite samples) also exhibit prolate spheroidal distributions of the sample
mean polarization vector with cylindrical symmetry about the
population mean polarization vector.  However, in this case, the axial
ratio is inconsistent with the degree of polarization of the
population mean Stokes parameters.

3. In the case of disjoint modes, the standard deviation of the
primary polarization $\varsigma_1$ may exceed the standard deviation
of the total intensity $\varsigma_0$.

However, several important phenomena invalidate the above simple observations.
First, as described in more detail in \Sec{noise}, the contribution
of superposed noise from both the sky and the instrument must be
properly subtracted before the covariances between the Stokes
parameters that are intrinsic to the source can be interpreted.
Second, the equations that describe superposed and composite
samples are valid only when the modes are statistically independent.
Any dependence between the modes, such as partial coherence or
covariant intensities, will invalidate points 1 and 3 above.
Partial coherence of orthogonally polarized modes
\citep[e.g.][]{gan97} is a plausible means of generating
non-orthogonal modes and, in the case of composite samples, covariant mode
intensities cause the variance of the primary polarization to differ
from that of the total intensity \citep[e.g.][]{ms98}.
In particular, anticorrelated mode intensities can mimic the
observable effects of disjoint modes by causing the variance of the
primary polarization to be greater than that of the total intensity.
Third, the effects of propagation through inhomogeneities in the
electron density \citep[e.g.][]{cbh+04} and magnetic field
\citep[e.g.][]{mm98} of the interstellar medium have not been
considered.  
Finally, as described in more detail in \Sec{modulation}, amplitude
modulation alters the form of the covariance matrix in a manner that
depends on the correlated structure of the amplitude modulating
function.
To distinguish between pulsar-intrinsic and scintillation-induced
variability, future work will incorporate analyses based on the
autocorrelation functions of the Stokes parameters
\cite[e.g.][]{ch77,cbh+04} and may employ techniques similar to the
fluctuation spectral analysis introduced by \citet{es03,edw04,es04}.
%


\subsection{Superposed Sky and Instrumental Noise}
\label{sec:noise}


Consider the observation of a source of electromagnetic radiation
described by the population mean Stokes parameters $S_\mathrm{S}$ and
covariance matrix ${\bf C}_\mathrm{S}$ superposed with statistically
independent sky and instrumental noise described by population mean
Stokes parameters $S_\mathrm{N}$ and covariance matrix ${\bf
  C}_\mathrm{N}$.
Section 4.3 of \cite{van09} incorrectly asserts that the incoherent
addition of unpolarized noise adds a constant term to each element of
the diagonal of the observed covariance matrix.
Rather, using \Eqn{covariance-superposed}, the
covariance matrix of the observed superposition is given by
\begin{equation}
{\bf C}_{\rm obs} = {\bf C}_\mathrm{S} + {\bf C}_\mathrm{N} + \outerSymm{S_\mathrm{S}}{S_\mathrm{N}}.
\label{eqn:covariance-noise}
\end{equation}
The above equation requires no assumptions regarding the distribution of
the electric field or the nature of the source (e.g. mode switching).
It is more generally applicable than Equation (9) of \citet{ch77},
which is valid only under the assumption that the system and sky noise
are unpolarized and normally distributed.
%
%
To solve \Eqn{covariance-noise} for the source-intrinsic covariance
matrix ${\bf C}_\mathrm{S}$, it is necessary to subtract both ${\bf
  C}_\mathrm{N}$ and $\outerSymm{S_\mathrm{S}}{S_\mathrm{N}}$ from the
observed covariance matrix.
In observations of radio pulsars, both $S_\mathrm{N}$ and ${\bf
  C}_\mathrm{N}$ are readily obtained from the off-pulse noise
statistics.
However, if it can be assumed that the electric field of the noise is
normally distributed, then greater sensitivity is achieved by
estimating the covariances between the noise Stokes parameters via
${\bf C}_\mathrm{N} = \spinorBilinear{S_\mathrm{N}}{S_\mathrm{N}}$.
The Stokes parameters of the source $S_\mathrm{S}$ are obtained by
subtracting $S_\mathrm{N}$ from the observed Stokes parameters $S$.
Given this information, \Eqn{covariance-noise} is trivial to solve for
${\bf C}_\mathrm{S}$.


\subsection{Amplitude Modulation}
\label{sec:modulation}


The derivation of the covariances between the sample mean Stokes
parameters presented in Section 3.3 of \citet{van09} begins with the
incorrect assertion that, for unpolarized radiation, the covariance
matrix {\bf C} is proportional to the $4\times4$ identity matrix,
regardless of the distribution of the electric field vector.
This assertion derives from the erroneous presumption that
uncorrelated signals are also statistically independent.\footnote
{ The faulty reasoning proceeds as follows.  First, because the
  radiation is unpolarized, the instantaneous intensities of the
  uncorrelated electric field vector components must also be
  uncorrelated; therefore, because $S_0$ and $S_1$ are the sums and
  differences of uncorrelated random values, their variances should be
  equal.
  Second, the statistics of unpolarized radiation should be
  independent of the basis in which it is measured; therefore, the
  variances of all three components of the Stokes polarization vector
  should be equal.
  A simple counter-example disproves both of these assumptions.
  Consider an electric field vector with components that have random,
  uncorrelated phases but equal amplitudes given by the random variate
  $\sqrt{a}$.
  The resulting signal is completely unpolarized; however, the
  field component intensities are completely correlated.
  In this case, the variance of $S_0$ is four times the variance of
  $a$, the variance of $S_1$ is zero, and the variances of $S_2$ and
  $S_3$ are equal to the second moment of $a$.  }

\noindent
This flawed reasoning is also reflected in Section 4.1 of
\citet{van09}, in which it is incorrectly argued that scalar amplitude
modulation uniformly scales the covariances of the Stokes parameters
by reducing the effective statistical degrees of freedom.

In fact, scalar amplitude modulation introduces statistical dependence
between the components of the electric field vector, such that their
instantaneous intensities are correlated.
As the total intensity is the sum of the intensities of the two field
components, positive covariance between the field component
intensities increases the variance of the amplitude-modulated total
intensity, even in the case of unpolarized radiation.
In contrast, each of the components of the Stokes polarization vector 
can be represented as differences in the intensities of the electric field
components in a given basis \citep{bw70}.
%
%
Therefore, positive covariance between the field component
intensities decreases the variances of the amplitude-modulated Stokes
polarization vector components.
That is, when the electric field vector is amplitude modulated by a
scalar multiplier, the variances of the components of the Stokes
polarization vector are expected to be less than the variance of the
total intensity.

The above reasoning is formally demonstrated by considering
amplitude modulation of the electric field vector by a statistically
independent random and real-valued dimensionless scalar $\sqrt{u}$,
producing $\mbf{e}^\prime = \sqrt{u}\mbf{e}$.
Using a variation of the technique introduced by \citet{goo60}, the
covariances between the modulated instantaneous Stokes parameters are
shown to be
%
%
%
%
\begin{equation}
{\bf C}^\prime = \left( \varsigma_u^2 + 1\right) {\bf C}
                 + \varsigma_u^2\, S\otimes S,
\label{eqn:covariance-modulation}
\end{equation}
where $\varsigma_u^2$ is the variance of $u$,
{\bf C} is the covariance matrix of the unmodulated instantaneous
Stokes parameters, $S$ is the population mean Stokes four-vector, and
without any loss of generality it is assumed that scalar amplitude
modulation does not alter the population mean Stokes parameters
(i.e.\ $\langle u \rangle = 1$).
In the natural basis where $S_0 \ge S_1 \ge 0$ and $S_2=S_3=0$, it is
readily seen that the last term in the right-hand side of
\Eqn{covariance-modulation} increases only the variances of the total
intensity $S_0$ and the primary polarization $S_1$ and the covariance
between them.
Furthermore, for partially polarized radiation where $S_0 > S_1$,
amplitude modulation increases the variance of the total intensity
$\varsigma_0^2$ by more than it increases the variance of the
primary polarization $\varsigma_1^2$.
This provides a simpler explanation for the observation made in
Section 4.4 of \citet{van09}, in which it is argued that orthogonally
polarized modes with covariant intensities could explain $\varsigma_0
> \varsigma_1$.
Rather, amplitude modulation causes $\varsigma_0 > \varsigma_1$ even
when only a single mode of emission contributes to the observed
signal.

\Eqn{covariance-modulation} is valid only for the instantaneous Stokes
parameters.
To derive the covariances between the sample mean Stokes parameters,
any correlations between instances of the amplitude modulating scalar
variate (e.g.\ subpulse structure) must be considered, which is beyond
the scope of this paper.
Temporally correlated structure of the amplitude modulation function
has been rigorously studied in the seminal works of \citet{ric75} and
\citet{cor76}.


\section{Conclusion}
\label{sec:discussion}

The statistical framework presented in this paper can be applied to
sources with arbitrary distributions of flux density; it also
incorporates the effects of instrumental integration over finite
samples.  Therefore, it is ideally suited to the study of
variability in polarized radiation on short timescales, such as the
polarization of subpulse structure in radio pulsar
emission.

Consideration of integration over finite samples highlights the
following important results.  First, the arguments that have been
presented to date in support of superposed modes of pulsar radiation
apply equally well to a composite sample of unresolved disjoint modes.
For example, both superposition and composition of modes result in
depolarization when integrated over a finite sample.
As argued in the introduction and elaborated in \App{gxv+99},
observational evidence of unresolved disjoint modes has already been
presented in the published literature.
In \App{comparison_with_crb78}, the statistical test defined by
Equation (5) of CRB is shown to be valid only when the orthogonally
polarized modes are assumed to be 100\% polarized and only when the
second moments of the {\emph{instantaneous} Stokes parameters are
  measured.
When this test is applied to single-pulse data comprised of sample
mean Stokes parameters, a composite sample of mutually exclusive modes
will be incorrectly identified as superposed.

In principle, it is possible to differentiate between the three
regimes of mode combination (disjoint, superposed, and composite
samples) through analysis of the covariances between all four Stokes
parameters.
However, various physical phenomena complicate interpretation.
Most importantly, amplitude modulation -- primarily that intrinsic to
the pulsar emission mechanism, but also that arising in the
interstellar medium -- significantly alters the structure of the
$4\times4$ covariance matrix.
The impact of amplitude modulation on the covariances between the
sample mean Stokes parameters depends on the correlated structure of
the amplitude modulating function (e.g.\ subpulse structure).
More detailed treatments of phenomena such as amplitude modulation
\citep[e.g.][]{ovdb13}, partial mode coherence \citep[e.g.][]{gan97},
covariant mode intensities \citep[e.g.][]{ms98}, and scintillation in
a magnetized plasma \citep[e.g.][]{mel93a,mel93b} are required before
the statistical framework presented in this paper will be sufficiently
developed to interpret observational data.
Therefore, a set of simulations were developed to verify the equations
presented in this work; these are described in \App{simulations}.

\acknowledgements

Parts of this research were supported by the Australian Research
Council Centre of Excellence for All-sky Astrophysics (CAASTRO),
through project number CE110001020, and the Australian Laureate
Fellowships scheme, through project number FL150100148.
The authors are grateful to Pablo Rosado for helping to test the
derivation presented in \App{linear_to_Mueller}.  We also thank
Damien Hicks for useful discussions and feedback on the paper.

\begin{appendix}

\setcounter{equation}{48}
\def\theequation{\arabic{equation}}


\section{Evidence for disjoint modes observed by Gangadhara et al. (1999)}
\label{app:gxv+99}

The scatter plots presented in Figure 3 of \citet{gxv+99}, hereafter
G+99, demonstrate that the degree of polarization of single pulses
(both fractional linear and fractional circular polarization)
increases as the instrumental integration length (or sample size) is
decreased.
To support the argument that this correlation represents plausible
evidence of unresolved, disjoint, and orthogonally polarized modes, it
is necessary to rule out a comparably plausible alternative
interpretation:
the observed correlation between temporal resolution and degree of
polarization could be a manifestation of a well known statistical
bias to the degree of polarization that increases as the sample size
is decreased; e.g.\ see Figure 3 of \citet{van09}, hereafter vS09.

G+99 observed PSR~B1133+16 using an instrument with a
bandwidth of $\Delta\nu=40$~MHz and, even on the shortest time scale
presented, $\tau_\mathrm{min}=150~\mu$s, the sample size (given by
the time-bandwidth product) is
$n=\tau_\mathrm{min}\Delta\nu=6\times10^3$.
For normally distributed noise, Equation (B4) of vS09 predicts a
maximum expected bias to the degree of polarization of less than 2\%,
which is too small to explain the correlation observed by G+99.
However, as discussed in Section 4 of vS09, the effective sample size
(or statistical degrees of freedom) may be reduced
by amplitude modulation and wave coherence, such as that introduced by
multipath propagation in the interstellar medium.
The scintillation bandwidth of PSR~B1133+16 is around 60~MHz at 1~GHz
\citep{cor86} and the corresponding coherence time in the observations
by G+99 (made at a center frequency of 1.41~GHz) is 4 to 5 orders of
magnitude smaller than $\tau_\mathrm{min}$.  Therefore, the impact of
scattering on the effective sample size is negligible.

To estimate the order of magnitude of the bias to the degree of
polarization induced by amplitude modulation, it is necessary to
characterize and model the physical nature of subpulse structure,
which varies between pulsars.
For the purposes of this argument, it suffices to describe the
subpulse structure by its characteristic width $\tau$ and the
modulation index $\beta$ that is observed after the instrument
integrates over some number of unresolved subpulses.
Conservative selection of these parameters for PSR~B1133+16 is based
on the following derivation of the relationship between them.

Consider normally distributed unpolarized noise that is modulated by a
contiguous sequence of rectangular subpulses.
Assume that the unpolarized noise has a population mean total
intensity of unity and that the characteristic subpulse width $\tau$
corresponds to a sub-sample size $n^\prime$; therefore, before modulation,
the sub-sample mean total intensity $\bar{S}^\prime_0$ has a variance of
$(2n^\prime)^{-1}$, as given by \Eqn{natural_covariance}.
As noted in the introduction, radio pulsar emission
typically exhibits log-normal distributions of subpulse amplitude
\citep[e.g.][]{cjd03a,ovb+14}.
Therefore, let the amplitude $u$ of each rectangular subpulse be
drawn from a log-normal distribution; i.e.\ $u=\exp(v)$, where $v$ is
a normally distributed random variate with zero mean and standard
deviation $\varsigma$, such that the mean and variance of $u$ are
$\langle u \rangle = \exp\left(\varsigma^2/2\right)$
and
$\varsigma_u^2 = \exp\left(2\varsigma^2\right)-\exp\left(\varsigma^2\right)$
respectively.  

Now consider the average of $N^\prime=n/n^\prime$ such subpulses, yielding the
sample mean of $N^\prime$ statistically independent instances of the
amplitude modulated sub-sample mean intensity, $u\bar{S}_0^\prime$.  The
variance of the sample mean normalized by the square of the population
mean yields the square of the modulation index
\begin{equation}
\beta^2 = {1\over n}\left[\left(\exp\left(\varsigma^2\right)-1\right)
                    \left(n^\prime+\frac{1}{2}\right)
                    +\frac{1}{2}\right],
\label{eqn:rectangular_modulation}
\end{equation}
which is trivially solved for the value of $\varsigma$ required to
yield a chosen modulation index, given the sample size $n$ and the
sub-sample size $n^\prime$ of the modulating subpulses.

For a fixed sample size, \Eqn{rectangular_modulation} shows that the modulation
index decreases as the subpulse width decreases, which necessitates a
larger value of $\varsigma$ to achieve a certain sample mean intensity
modulation index.
In turn, this decreases the effective sample size and increases the
bias to the degree of polarization.
Therefore, to present a conservative upper limit on
bias, the shortest microstructure characteristic timescale,
$\tau_\mathrm{\mu-narrow}\sim 10~\mu$s, estimated by \cite{pbc+02} is adopted.
Similarly, from the phase-resolved modulation index of PSR~B1133+16
presented in Figure A.19 of \cite{wes06}, the maximum value of
$\beta\sim2.4$ is assumed.
The adopted values yield $n^\prime=\tau_\mathrm{\mu-narrow}\Delta\nu=400$,
$N^\prime=\tau_\mathrm{min}/\tau_\mathrm{\mu-narrow}=15$,
and $\varsigma\sim2.1$;
given these parameters, a simple simulation (described in
\App{simulations}) shows that the bias to the degree of polarization is
less than 4\%.
%
%
%
Although the subpulse structure of pulsars does not consist of a
sequence of rectangular impulses, this tractable basic model is
sufficient to demonstrate to first order that amplitude modulation by
narrow subpulses with log-normally distributed amplitudes does not
significantly reduce the statistical degrees of freedom in the pulsar
signal.
Therefore, as neither wave coherence nor amplitude modulation reproduce the
observed correlation between degree of polarization and temporal
resolution, it is unlikely that the correlation can be dismissed as a
consequence of the bias associated with small number statistics.


\section{Comparison with Cardoso (1991)}
\label{app:comparison_with_car91}

\Eqns{indexOuterBilinear}{through}{double_contraction} incorporate a
couple of conventions that are different to those used by
\cite{car91}.
Although these differences are of no consequence in this work, they
deserve mention.
First, note that Equation (6) of \cite{car91} includes the Hermitian
transpose of the second operand {\bf B} in the definitions of the
tensor products, $\otimes$ and \stimes.
To maintain consistency with the notation of
\Eqn{nth_moment}, where the tensor product and adjoint operator are
separately and explicitly invoked, the implicit adjoint operator is omitted
from the definitions in \Eqns{indexOuterBilinear}{and}{indexSpinorBilinear}
and therefore does not appear on the right hand sides of
\Eqns{outerBilinear}{and}{spinorBilinear}.
This distinction is inconsequential because in this work $\mbf\rho$,
{\bf A} and {\bf B} are self-adjoint and linearly related to
real-valued Stokes parameters.

Also note that the double contraction defined in
\Eqn{double_contraction} uses an index notation convention that is
different to the one that appears on the left hand sides of Equation
(6) in \cite{car91}.
As inferred from the indeces on the tensor coordinates in the text
that follows Equation (6), the convention used by \cite{car91} is
defined by transposing $\mbf\rho$ in \Eqn{double_contraction}.
Again, this difference is of no consequence because the tensor
coordinates are of secondary importance to the transformation
properties of \outerBilinear{\bf A}{\bf B} and \spinorBilinear{\bf
  A}{\bf B}.

More important than these minor differences is the fundamental
consistency between the findings of \cite{car91} and the results of
\Sec{Stokes_statistics}.  After accounting for the difference in index
notation convention mentioned above, the third symmetry identified by
\citet{car91} is easily seen to be equivalent to the transpose over
contravariant tensor indeces that converts \outerBilinear{\bf A}{\bf
  B} into \spinorBilinear{\bf A}{\bf B} (and vice versa).
This leads to a useful interpretation of the ``Rank One Lemma''
demonstrated by \citet{car91}.
In the special case of 100\% polarized radiation, the coherency matrix
has a rank of one because it has only one non-zero eigenvalue
\citep[e.g.\ Eq.~{[7]} of][and the discussion that follows]{van09}.
In this case, the Lorentz invariant of the associated Stokes
parameters $S^2=0$ and $\outerBilinear{S}{S}=\spinorBilinear{S}{S}$;
that is, in the case of 100\% polarization, the two tensor products are
invariant under the third symmetry and equal to each other.
Note that the instantaneous coherency matrix is always rank one and
the instantaneous Stokes parameters always satisfy
$\outerBilinear{s}{s}=\spinorBilinear{s}{s}$.


\newcommand{\EqnSpinorStokes}{\ref{eqn:spinorStokes}}

\section{Derivation of Equation (\protect\EqnSpinorStokes)}
\label{app:linear_to_Mueller}

To derive \Eqn{spinorStokes}, substitute
${\bf A}=A_\alpha\pauli{\alpha}/2$ and ${\bf B}=B_\beta\pauli{\beta}/2$ 
into \Eqn{spinorMueller} to express it in terms of the associated
Stokes parameters,
\begin{equation}
\left\{\spinorMueller{A}{B}\right\}_\irow^\icol
  = \frac{1}{2} \dc{\pauli{\irow}}{\left({\bf A}\,\pauli{\icol}\,{\bf B}\right)}
  = \frac{1}{8} A_\alpha B_\beta \tr{\pauli{\irow}\,\pauli{\alpha}\,\pauli{\icol}\,\pauli{\beta}}.
\label{eqn:commutator_Mueller}
\end{equation}
For each of the 16 elements of \spinorMueller{A}{B}, a total of 16
terms arise in the above double sum over $\alpha$ and $\beta$.
The trace of a product of four Pauli matrices has been derived in
Appendix B of \citet{mel93a}.  To arrive at a more compact,
coordinate-free representation, the $4\times4$ matrix is partitioned
into the following four parts.
\begin{enumerate}
\item $\icol=\irow=0$
\item $\icol=0$ and $\irow=\srow>0$
\item $\icol=\scol>0$ and $\irow=0$
\item $\icol=\scol>0$ and $\irow=\srow>0$
\end{enumerate}
For each part, the double sum is solved as follows.

\begin{enumerate}
\item When $\irow=\icol=0$,
  the result is simply one quarter of the Euclidean inner product of
  the Stokes four-vectors; i.e.,
\begin{equation}
\label{eqn:M00}
\left\{\spinorMueller{A}{B}\right\}_0^0 = \frac{1}{2} \tr{\bf AB}
     = \frac{1}{4} A_\irow B_\irow.
\end{equation}
\item When $\irow=\srow>0$ and $\icol=0$,
the double sum can be partitioned into the same four parts
\begin{enumerate}
\item $\alpha=\beta=0$: \tr{\pauli{\srow}}=0
\item $\alpha=0$ and $\beta=b>0$: $\tr{\pauli{\srow}\pauli{b}}=2\delta_{\srow b}$
\item $\alpha=a>0$ and $\beta=0$: $\tr{\pauli{\srow}\pauli{a}}=2\delta_{\srow a}$
\item $\alpha=a>0$ and $\beta=b>0$: $\tr{\pauli{\srow}\pauli{a}\pauli{b}}=2i\epsilon_{\srow ab}$ 
\end{enumerate}
where $\epsilon_{jkl}$ is the rank 3 permutation pseudotensor, yielding
\begin{equation}
\label{eqn:Mk0}
\left\{\spinorMueller{A}{B}\right\}_\srow^0 
     = \frac{1}{4} \left( A_0 B_\srow + A_\srow B_0 + i \epsilon_{\srow ab} A_a B_b \right).
\end{equation}
\item Similarly, when $\irow=0$ and $\icol=\scol>0$,
\begin{equation}
\label{eqn:M0j}
\left\{\spinorMueller{A}{B}\right\}_0^\scol 
     = \frac{1}{4} \left( A_\scol B_0 + A_0 B_\scol + i \epsilon_{a\scol b} A_a B_b \right)
\end{equation}
\item When $\irow=\srow>0$ and $\icol=\scol>0$, partition the double sum to yield
\begin{enumerate}
\item $\alpha=\beta=0$: $\tr{\pauli{\srow}\pauli{\scol}}=2\delta_{\scol\srow}$
\item $\alpha=0$ and $\beta=b>0$: $\tr{\pauli{\srow}\pauli{\scol}\pauli{b}}=2i\epsilon_{\srow\scol b}$ 
\item $\alpha=a>0$ and $\beta=0$: $\tr{\pauli{\srow}\pauli{a}\pauli{\scol}}=2i\epsilon_{\srow a\scol}$ 
\item $\alpha=a>0$ and $\beta=b>0$: $\tr{\pauli{\srow}\pauli{a}\pauli{\scol}\pauli{b}}
= 2\left( \delta_{\srow a}\delta_{\scol b} - \delta_{\srow\scol}\delta_{ab} + \delta_{\srow b}\delta_{a\scol} \right)$
\end{enumerate}
and
\begin{equation}
\label{eqn:Mkj}
\left\{\spinorMueller{A}{B}\right\}_\srow^\scol = \frac{1}{4} \left( 
  A_\srow B_\scol + A_\scol B_\srow + \delta_{\scol\srow} A \cdot B
+ i\epsilon_{\srow\scol b}A_0B_b
+ i\epsilon_{\srow a\scol}A_aB_0 \right).
\end{equation}
\end{enumerate}
Noting that $A_\irow B_\irow = 2A_0 B_0 - A \cdot B$,
\Eqns{M00}{through}{Mkj} can be combined to produce
\begin{equation}
\spinorMueller{A}{B} = \frac{1}{4} 
 \left( A \otimes B + B \otimes A
 - {\mbf\eta}\Linner{A}{B} + i A \wedge B \right).
\label{eqn:appSpinorMueller}
\end{equation}
Here, $A \wedge B$ is the covariant exterior product of $A$ and $B$, a
$4\times4$ antisymmetric matrix
(i.e.\ $\exterior{A}{B}=-\exterior{B}{A}$) defined by
\begin{equation}
\left\{A \wedge B\right\}_\irow^\icol
\equiv \epsilon_{\irow\alpha\icol\beta} A^\alpha B^\beta,
\label{eqn:exterior}
\end{equation}
where $\epsilon_{\alpha\beta\delta\gamma}$ is the rank 4 permutation
pseudotensor, $A^\alpha=\eta^{\alpha\gamma}A_\gamma$ and
$B^\beta=\eta^{\beta\gamma}B_\gamma$.
Finally, as in \Eqn{outerStokes}, define
$\spinorBilinear{A}{B}=2\spinorMueller{A}{B}$.

\section{Comparison with Cordes et al.\ (1978)}
\label{app:comparison_with_crb78}

To distinguish between disjoint and superposed modes, Section III.~c)
of CRB proposes a statistical test that starts by describing the
linear polarization $L=Q+iU$ at a given pulse longitude as a linear
combination of orthogonally polarized modes,
$L=L_1+L_2=\left(|L_1|-|L_2|\right)\exp(2i\psi)$.
It is asserted that intermediate position angles between $\psi$ and
$\psi+\pi/2$ do not occur; therefore, without any loss of generality,
choose $\psi=0$ and define the orthogonally polarized modes such that
the instantaneous Stokes parameters $a=2\left[I_1,|L_1|,0,0\right]$
and $b=2\left[I_2,-|L_2|,0,0\right]$.
Note that, because $a_2=a_3=b_2=b_3=0$, the standard deviations
$\varsigma_2=\varsigma_3=0$ and, referring to \Eqn{natural_covariance},
this implies that the Lorentz invariant $S^2=0$.
That is, the modes defined in Section III.~c) of CRB are implicitly
assumed to be 100\% polarized.
In this case, the mode intensities, $I_1=|L_1|$ and $I_2=|L_2|$;
however, to facilitate comparison with CRB, $I_1$, $I_2$, $|L_1|$, and
$|L_2|$ are treated as distinct random variates until these two
equalities are required.

CRB also assume that both modes occur equally frequently such that,
when the modes are disjoint, the population mean Stokes parameters are
$S=(A+B)/2=\left[\mean{I}, \mean{L}, 0, 0\right]$,
where
$A=\mean{a}$ and $B=\mean{b}$.
After substitution of the above definitions and $n=1$ and $F=0.5$ into
\Eqn{covariance-disjoint}, rearrange and solve for the moments of the
instantaneous Stokes parameters,
\begin{eqnarray}
\mean{s_\irow s_\icol} = \Celement{{\bf C}_d} + \mean{s_\irow}\mean{s_\icol}
  & = & \frac{1}{2}\left( \Celement{{\bf C}_A} + \Celement{{\bf C}_B} \right)
      + \frac{1}{4}\left[ \left(A_\irow - B_\irow\right)\left(A_\icol - B_\icol\right)
                        + \left(A_\irow + B_\irow\right)\left(A_\icol + B_\icol\right)
                  \right]
\nonumber \\
  & = & \frac{1}{2}\left( \mean{a_\irow a_\icol} + \mean{b_\irow b_\icol} \right).
\end{eqnarray}
The above equation yields 
$\mean{L^2}=\mean{s_1^2}=2\left(\mean{L_1^2}+\mean{L_2^2}\right)$ 
and, noting that $\mean{I_1 I_2}=0$ for mutually exclusive modes,
$\mean{I^2}=\mean{s_0^2}=2 \mean{\left(I_1+I_2\right)^2}$.  
Therefore, the covariances between the instantaneous Stokes parameters of
disjoint modes described in \Sec{disjoint} are consistent with
disjoint modes in Equation (5) of CRB.  Furthermore, although not
explicitly noted in CRB, because the modes
are implicitly assumed to be 100\% polarized,
$\mean{d_L^2} \equiv \mean{L^2}/\mean{I^2} = 1$ in the case of
disjoint modes.

For superposed modes, divide $a$ and $b$ by 2,
then substitute $n=1$ and $S=A+B$ into \Eqn{covariance-superposed} and
solve for the moments,
\begin{eqnarray}
\mean{s_\irow s_\icol} = \Celement{{\bf C}_s} + \mean{s_\irow}\mean{s_\icol}
 & = & \Celement{{\bf C}_A} + \Celement{{\bf C}_B} + \Celement{\outerSymm{A}{B}} \\
& = & \mean{a_\irow a_\icol} + \mean{b_\irow b_\icol} + 2 A_\irow B_\icol + 2 A_\icol B_\irow - \eta_\irow^\icol \Linner{A}{B}.
\end{eqnarray}
Owing to the assumption that the modes are 100\% polarized,
$\Linner{A}{B}=\mean{I_1}\mean{I_2} + \mean{|L_1|}\mean{|L_2|} = 2 \mean{|L_1|}\mean{|L_2|}$,
and the above equation yields
$\mean{L^2}=\mean{s_1^2}=\mean{L_1^2}+\mean{L_2^2}-2\mean{|L_1|}\mean{|L_2|}$.
Furthermore, because the mode intensities are uncorrelated,
$\langle I_1 I_2 \rangle = \langle I_1 \rangle\langle I_2 \rangle$ and
$\mean{I^2}=\mean{s_0^2}=\mean{\left(I_1+I_2\right)^2}$.
Therefore, the covariances between the instantaneous Stokes parameters of
superposed modes as described in \Sec{superposed} are consistent with 
superposition of modes in Equation (5) of CRB only if the modes
are assumed to be 100\% polarized.

To arrive at the above results for disjoint and superposed modes, it
is necessary to consider the instantaneous Stokes parameters; that is,
to be consistent with CRB, the Stokes sample size must be unity
($n=1$).
Under this assumption, mutually exclusive modes are disjoint by
definition.
To consider a Stokes sample composed of mutually exclusive and
orthogonally polarized instances, replace the instantaneous Stokes
parameters $a$ and $b$ with the sub-sample mean Stokes parameters
$\bar{A}^\prime=2\left[I_1,|L_1|,0,0\right]$ and
$\bar{B}^\prime=2\left[I_2,-|L_2|,0,0\right]$, formed after respectively
averaging over all instances in mode $A$ and all instances in mode
$B$.  In this case, the matrix of covariances between
the sample mean Stokes parameters is given by
%
%
%
%
%
%
\begin{equation}
\bar{\bf C}_c = f^2 \bar{\bf C}_A^\prime + (1-f)^2 \bar{\bf C}_B^\prime,
\label{eqn:covariance-composite-subsample}
\end{equation}
where 
$\bar{\bf C}_A^\prime={\bf C}_A(fn)^{-1}$ 
and 
$\bar{\bf C}_B^\prime={\bf C}_B\left[(1-f)n\right]^{-1}$
are the covariances between the sub-sample mean Stokes parameters
after averaging over $fn$ instances in mode $A$ and $(1-f)n$ instances
in mode $B$, respectively.
Substitute $f=0.5$ into the above equation and rearrange to yield the
moments of the composite sample mean Stokes parameters,
\begin{eqnarray}
\mean{\bar{S}_\irow \bar{S}_\icol} = \Celement{\bar{\bf C}_c}
                              + \mean{\bar{S}_\irow}\mean{\bar{S}_\icol}
 & = & \frac{1}{4}\left( \Celement{\bar{\bf C}_A^\prime}
                   + \Celement{\bar{\bf C}_B^\prime} \right)
     + \frac{1}{4}\left(A_\irow + B_\irow\right)\left(A_\icol + B_\icol\right)
\nonumber \\
& = & \frac{1}{4}\left( \mean{\bar{A}_\irow^\prime \bar{A}_\icol^\prime}
+ \mean{\bar{B}_\irow^\prime \bar{B}_\icol^\prime}
+ A_\irow B_\icol + A_\icol B_\irow \right).
\end{eqnarray}
The above equation yields
$\mean{L^2}=\mean{\bar{S}_1^2}=\mean{L_1^2}+\mean{L_2^2}-2\mean{|L_1|}\mean{|L_2|}$
and
$\mean{I^2}=\mean{\bar{S}_0^2}=\mean{\left(I_1+I_2\right)^2}$.
Therefore, the second moments of the composite sample mean Stokes
parameters are also consistent with the superposition of modes in
Equation (5) of CRB.  That is, given only the second moments of the
sample mean total and linearly polarized intensities, it is not
possible to distinguish between superposed modes and Stokes samples
that are composed of mutually exclusive states.

%
%


\section{Verification by Simulation}
\label{app:simulations}

To verify the equations presented in this paper, the following Monte
Carlo simulation was repeatedly performed over a wide range of input
parameters and conditions.

\begin{enumerate}
\item Generate a sequence of $M$ random electric field vector
  instances $\mbf{e}$, each with statistically independent and
  identically distributed (iid) circular complex normal components;
  such a sequence is described by the population mean Stokes
  parameters [1,0,0,0].
  
\item To yield the desired population mean Stokes parameters, $S_\irow$,
  transform each electric field vector instance by the Hermitian square
  root of $2\mbf\rho=S_\irow\,\pauli{\irow}$.

\item Optionally perform amplitude modulation by multiplying each
  instance of $\mbf{e}$ by an iid random variate $u$ that is drawn
  from a log-normal distribution.  The log-normally distributed
  variate is generated from a normally distributed iid variate with
  zero mean and standard deviation $\varsigma$ and is normalized by
  the mean of the distribution, $\langle u \rangle =
  \exp(\varsigma^2/2)$, such that the mean of the amplitude modulating
  function is unity.
  To simulate rectangular subpulses defined by the sub-sample size $n$, as
  described in \App{gxv+99}, a single value of $u$ is applied to $n^\prime$
  consecutive instances of $\mbf{e}$.

\item If simulating \emph{superposed samples}, repeat all of the previous
  steps to produce $M$ instances of the electric field vector in the
  other mode then, for each instance of the electric field vectors
  from modes $A$ and $B$, produce $M$ new instances
  $\mbf{e}=\mbf{e}_A+\mbf{e}_B$.
  
\item Compute the instantaneous Stokes parameters,
  $s_\irow=\mbf{e}^\dagger\pauli{\irow}\mbf{e}$.

\item Optionally divide the sequence of $M$ instantaneous Stokes
  vectors into mutually exclusive Stokes samples of $n$ instances,
  yielding a sequence of $N=M/n$ Stokes samples.  This step is not
  optional when simulating composite samples.

\item If simulating \emph{composite samples}, replace $(1-f)n$ instances
  in each Stokes sample with instantaneous Stokes vectors in the
  other mode.

\item If simulating \emph{disjoint samples}, replace $(1-F)N$ Stokes
  samples with Stokes samples that contain only instantaneous
  Stokes vectors in the other mode.

\item For each Stokes sample, compute the sample mean Stokes
  parameters $\bar{S}_\irow$.

\item Compute the $4\times4$ covariances between the Stokes parameters
  using either the $M$ instantaneous Stokes parameters or the $N$
  sample mean Stokes parameters.  Verify that the computed covariance
  matrix matches the theoretical prediction within the uncertainty due
  to noise.
  
\end{enumerate}

The above simulation is implemented in C++ and is freely available as
the {\sc epsic} open source software package for simulating the polarization
of electromagnetic radiation\footnote{See {\tt http://straten.github.io/epsic}}.

\end{appendix}

\bibliographystyle{aasjournal}
\bibliography{journals,modrefs,psrrefs,../local,crossrefs}

\begin{thebibliography}{}
\expandafter\ifx\csname natexlab\endcsname\relax\def\natexlab#1{#1}\fi

\bibitem[{Allen \& Melrose(1982)}]{am82}
Allen, M., \& Melrose, D. 1982, PASA, 4, 365

\bibitem[{Amblard {et~al.}(1996)Amblard, Gaeta, \& Lacoume}]{agl96a}
Amblard, P., Gaeta, M., \& Lacoume, J. 1996, Signal Processing, 53, 1

\bibitem[{Backer(1970)}]{bac70}
Backer, D.~C. 1970, Nature, 228, 42

\bibitem[{Backer \& Rankin(1980)}]{br80}
Backer, D.~C., \& Rankin, J.~M. 1980, ApJS, 42, 143

\bibitem[{{Barakat}(1963)}]{bar63}
{Barakat}, R. 1963, J. Opt. Soc. Am., 53, 317

\bibitem[{Born \& Wolf(1970)}]{bw70}
Born, M., \& Wolf, E. 1970, Principles of Optics: Electromagnetic Theory of
  Propagation, Interference and Diffraction of Light (New York: Pergamon)

\bibitem[{Bracewell(1986)}]{bra86b}
Bracewell, R. 1986, The Fourier Transform and its Applications (New York:
  McGraw-Hill)

\bibitem[{Britton(2000)}]{bri00}
Britton, M.~C. 2000, ApJ, 532, 1240

\bibitem[{Brosseau \& Barakat(1992)}]{bb92}
Brosseau, C., \& Barakat, R. 1992, Opt. Comm., 91, 408

\bibitem[{{Burns} \& {Clark}(1969)}]{bc69}
{Burns}, W.~R., \& {Clark}, B.~G. 1969, A\&A, 2, 280

\bibitem[{Cairns {et~al.}(2003)Cairns, Johnston, \& Das}]{cjd03a}
Cairns, I.~H., Johnston, S., \& Das, P. 2003, MNRAS, 343, 512

\bibitem[{Cardoso(1991)}]{car91}
Cardoso, J.-F. 1991, in ICASSP 91. 1991 International Conference on Acoustics,
  Speech, and Signal Processing (New York: IEEE), 3109 -- 3112

\bibitem[{Cloude(1986)}]{clo86}
Cloude, S. 1986, Optik, 75, 26

\bibitem[{Cognard {et~al.}(1996)Cognard, Shrauner, Taylor, \&
  Thorsett}]{cstt96}
Cognard, I., Shrauner, J.~A., Taylor, J.~H., \& Thorsett, S.~E. 1996, ApJ, 457,
  L81

\bibitem[{{Cordes}(1976)}]{cor76}
{Cordes}, J.~M. 1976, ApJ, 208, 944

\bibitem[{Cordes(1986)}]{cor86}
Cordes, J.~M. 1986, ApJ, 311, 183

\bibitem[{{Cordes} {et~al.}(2004){Cordes}, {Bhat}, {Hankins}, {McLaughlin}, \&
  {Kern}}]{cbh+04}
{Cordes}, J.~M., {Bhat}, N.~D.~R., {Hankins}, T.~H., {McLaughlin}, M.~A., \&
  {Kern}, J. 2004, ApJ, 612, 375

\bibitem[{Cordes \& Hankins(1977)}]{ch77}
Cordes, J.~M., \& Hankins, T.~H. 1977, ApJ, 218, 484

\bibitem[{Cordes {et~al.}(1978)Cordes, Rankin, \& Backer}]{crb78}
Cordes, J.~M., Rankin, J.~M., \& Backer, D.~C. 1978, ApJ, 223, 961

\bibitem[{{Cordes} {et~al.}(2006){Cordes}, {Rickett}, {Stinebring}, \&
  {Coles}}]{crsc06}
{Cordes}, J.~M., {Rickett}, B.~J., {Stinebring}, D.~R., \& {Coles}, W.~A. 2006,
  ApJ, 637, 346

\bibitem[{{Edwards}(2004)}]{edw04}
{Edwards}, R.~T. 2004, A\&A, 426, 677

\bibitem[{{Edwards} \& {Stappers}(2002)}]{es02}
{Edwards}, R.~T., \& {Stappers}, B.~W. 2002, A\&A, 393, 733

\bibitem[{{Edwards} \& {Stappers}(2003)}]{es03}
---. 2003, A\&A, 407, 273

\bibitem[{{Edwards} \& {Stappers}(2004)}]{es04}
---. 2004, A\&A, 421, 681

\bibitem[{{Ekers} \& {Moffet}(1969)}]{em69}
{Ekers}, R.~D., \& {Moffet}, A.~T. 1969, ApJ, 158, L1

\bibitem[{Eriksson {et~al.}(2010)Eriksson, Ollila, \& Koivunen}]{eok10}
Eriksson, J., Ollila, E., \& Koivunen, V. 2010, IEEE Transactions on Signal
  Processing, 58, 5400

\bibitem[{Gangadhara(1997)}]{gan97}
Gangadhara, R. 1997, A\&A, 327, 155

\bibitem[{{Gangadhara} {et~al.}(1999){Gangadhara}, {Xilouris}, {von
  Hoensbroech}, {Kramer}, {Jessner}, \& {Wielebinski}}]{gxv+99}
{Gangadhara}, R.~T., {Xilouris}, K.~M., {von Hoensbroech}, A., {et~al.} 1999,
  A\&A, 342, 474

\bibitem[{Gil {et~al.}(1991)Gil, Snakowski, \& Stinebring}]{gss91}
Gil, J.~A., Snakowski, J.~K., \& Stinebring, D.~R. 1991, A\&A, 242, 119

\bibitem[{Goodman(1960)}]{goo60}
Goodman, L.~A. 1960, Journal of the American Statistical Association, 55, 708

\bibitem[{Goodman(1963)}]{goo63}
Goodman, N.~R. 1963, Ann. of Math. Stat., 34, 152

\bibitem[{{Hankins} {et~al.}(2003){Hankins}, {Kern}, {Weatherall}, \&
  {Eilek}}]{hkwe03}
{Hankins}, T.~H., {Kern}, J.~S., {Weatherall}, J.~C., \& {Eilek}, J.~A. 2003,
  Nature, 422, 141

\bibitem[{Harding \& Tademaru(1981)}]{ht81}
Harding, A.~K., \& Tademaru, E. 1981, ApJ, 243, 597

\bibitem[{Heiles {et~al.}(1970)Heiles, Campbell, \& Rankin}]{hcr70}
Heiles, C., Campbell, D.~B., \& Rankin, J.~M. 1970, Nature, 226, 529

\bibitem[{{Jenet} \& {Gil}(2003)}]{jg03}
{Jenet}, F.~A., \& {Gil}, J. 2003, ApJ, 596, L215

\bibitem[{Karastergiou {et~al.}(2003)Karastergiou, Johnston, \& Kramer}]{kjk03}
Karastergiou, A., Johnston, S., \& Kramer, M. 2003, A\&A, 404, 325

\bibitem[{{Karastergiou} {et~al.}(2011){Karastergiou}, {Roberts}, {Johnston},
  {Lee}, {Weltevrede}, \& {Kramer}}]{krj+11}
{Karastergiou}, A., {Roberts}, S.~J., {Johnston}, S., {et~al.} 2011, MNRAS,
  415, 251

\bibitem[{Kendall {et~al.}(1987)Kendall, Stuart, \& Ord}]{kso87}
Kendall, M.~G., Stuart, A., \& Ord, J.~K. 1987, Kendall's Advanced Theory of
  Statistics (London: Charles Griffin)

\bibitem[{Kollo \& von Rosen(2005)}]{kr05}
Kollo, T., \& von Rosen, D. 2005, {Advanced Multivariate Statistics with
  Matrices} (Dordrecht: Springer)

\bibitem[{Krishnamohan \& Downs(1983)}]{kd83}
Krishnamohan, S., \& Downs, G.~S. 1983, ApJ, 265, 372

\bibitem[{{Liu} {et~al.}(2015){Liu}, {Karuppusamy}, {Lee}, {Stappers},
  {Kramer}, {Smits}, {Purver}, {Janssen}, \& {Perrodin}}]{lkl+15}
{Liu}, K., {Karuppusamy}, R., {Lee}, K.~J., {et~al.} 2015, MNRAS, 449, 1158

\bibitem[{Manchester {et~al.}(1975)Manchester, Taylor, \& Huguenin}]{mth75}
Manchester, R.~N., Taylor, J.~H., \& Huguenin, G.~R. 1975, ApJ, 196, 83

\bibitem[{McCullagh(1987)}]{mcc87}
McCullagh, P. 1987, Tensor Methods in Statistics, Monographs on Statistics and
  Applied Probability (London: Chapman and Hall)

\bibitem[{McKinnon \& Stinebring(1998)}]{ms98}
McKinnon, M., \& Stinebring, D. 1998, ApJ, 502, 883

\bibitem[{{McKinnon}(2002)}]{mck02}
{McKinnon}, M.~M. 2002, ApJ, 568, 302

\bibitem[{{McKinnon}(2003{\natexlab{a}})}]{mck03a}
---. 2003{\natexlab{a}}, ApJ, 590, 1026

\bibitem[{{McKinnon}(2003{\natexlab{b}})}]{mck03b}
---. 2003{\natexlab{b}}, ApJS, 148, 519

\bibitem[{{McKinnon}(2004)}]{mck04}
---. 2004, ApJ, 606, 1154

\bibitem[{{McKinnon}(2006)}]{mck06}
---. 2006, ApJ, 645, 551

\bibitem[{{McKinnon} \& {Stinebring}(2000)}]{ms00}
{McKinnon}, M.~M., \& {Stinebring}, D.~R. 2000, ApJ, 529, 435

\bibitem[{{Melrose}(1993{\natexlab{a}})}]{mel93a}
{Melrose}, D.~B. 1993{\natexlab{a}}, Journal of Plasma Physics, 50, 267

\bibitem[{{Melrose}(1993{\natexlab{b}})}]{mel93b}
---. 1993{\natexlab{b}}, Journal of Plasma Physics, 50, 283

\bibitem[{Melrose \& Macquart(1998)}]{mm98}
Melrose, D.~B., \& Macquart, J.-P. 1998, ApJ, 505, 921

\bibitem[{Mendel(1991)}]{men91}
Mendel, J. 1991, Proc. IEEE, 79, 278

\bibitem[{{Os{\l}owski} {et~al.}(2014){Os{\l}owski}, {van Straten}, {Bailes},
  {Jameson}, \& {Hobbs}}]{ovb+14}
{Os{\l}owski}, S., {van Straten}, W., {Bailes}, M., {Jameson}, A., \& {Hobbs},
  G. 2014, MNRAS, 441, 3148

\bibitem[{{Os{\l}owski} {et~al.}(2013){Os{\l}owski}, {van Straten}, {Demorest},
  \& {Bailes}}]{ovdb13}
{Os{\l}owski}, S., {van Straten}, W., {Demorest}, P., \& {Bailes}, M. 2013,
  MNRAS, 430, 416

\bibitem[{Picinbono(1994)}]{pic94}
Picinbono, B. 1994, Trans. Sig. Proc., 42, 3473

\bibitem[{{Popov} {et~al.}(2002){Popov}, {Bartel}, {Cannon}, {Novikov},
  {Kondratiev}, \& {Altunin}}]{pbc+02}
{Popov}, M.~V., {Bartel}, N., {Cannon}, W.~H., {et~al.} 2002, Astronomy
  Reports, 46, 206

\bibitem[{{Ransom} {et~al.}(2002){Ransom}, {Eikenberry}, \&
  {Middleditch}}]{rem02}
{Ransom}, S.~M., {Eikenberry}, S.~S., \& {Middleditch}, J. 2002, AJ, 124, 1788

\bibitem[{{Rickett}(1975)}]{ric75}
{Rickett}, B.~J. 1975, ApJ, 197, 185

\bibitem[{{Smirnov}(2011)}]{smi11}
{Smirnov}, O.~M. 2011, A\&A, 531, A159

\bibitem[{Stinebring {et~al.}(1984)Stinebring, Cordes, Rankin, Weisberg, \&
  Boriakoff}]{scr+84}
Stinebring, D.~R., Cordes, J.~M., Rankin, J.~M., Weisberg, J.~M., \& Boriakoff,
  V. 1984, ApJS, 55, 247

\bibitem[{{Stinebring} {et~al.}(2001){Stinebring}, {McLaughlin}, {Cordes},
  {Becker}, {Goodman}, {Kramer}, {Sheckard}, \& {Smith}}]{smc+01}
{Stinebring}, D.~R., {McLaughlin}, M.~A., {Cordes}, J.~M., {et~al.} 2001, ApJ,
  549, L97

\bibitem[{Sultan \& Tracy(1996)}]{st96}
Sultan, S.~A., \& Tracy, D.~S. 1996, Linear Algebra and its Applications,
  237/238, 191

\bibitem[{Taylor {et~al.}(1971)Taylor, Huguenin, Hirsch, \&
  Manchester}]{thhm71}
Taylor, J.~H., Huguenin, G.~R., Hirsch, R.~M., \& Manchester, R.~N. 1971,
  Astrophys. Lett., 9, 205

\bibitem[{Taylor {et~al.}(1975)Taylor, Manchester, \& Huguenin}]{tmh75}
Taylor, J.~H., Manchester, R.~N., \& Huguenin, G.~R. 1975, ApJ, 195, 513

\bibitem[{{van Straten}(2009)}]{van09}
{van Straten}, W. 2009, ApJ, 694, 1413

\bibitem[{{van Straten}(2010)}]{van10}
---. 2010, ApJ, 719, 985

\bibitem[{{Walker} {et~al.}(2004){Walker}, {Melrose}, {Stinebring}, \&
  {Zhang}}]{wmsz04}
{Walker}, M.~A., {Melrose}, D.~B., {Stinebring}, D.~R., \& {Zhang}, C.~M. 2004,
  MNRAS, 354, 43

\bibitem[{{Weltevrede} {et~al.}(2006){Weltevrede}, {Edwards}, \&
  {Stappers}}]{wes06}
{Weltevrede}, P., {Edwards}, R.~T., \& {Stappers}, B.~W. 2006, A\&A, 445, 243

\bibitem[{{Weltevrede} {et~al.}(2007){Weltevrede}, {Stappers}, \&
  {Edwards}}]{wse07}
{Weltevrede}, P., {Stappers}, B.~W., \& {Edwards}, R.~T. 2007, A\&A, 469, 607

\end{thebibliography}

\begin{appendix}

  The following additional material was not submitted to The Astrophysical
  Journal and was not reviewed by the referee.  It is provided as further
  information for the interested reader. \\ [5mm]
  
\setcounter{section}{6}
\setcounter{equation}{61}
\def\theequation{\arabic{equation}}

%
%

\subsection{Error in Section 5.2 of Amblard, Gaeta, and Lacoume (1996)}
\label{sec:error_in_agl96}

\noindent
In {\it Statistics for complex variables and signals - Part I: Variables},
Amblard, Gaeta, and Lacoume (1996) study the higher-order statistics of
multivariate complex random values using tensor notation.
At the bottom of the left-hand column of page 9, they assert that the
fourth-order cumulant $C_2^2$ of the vector $\mbf{Z}$ is related to
the fourth-order moment $M_2^2$ and second-order moment $M_1^1$ as
follows
\begin{equation}
C_2^2 = M_2^2 - 2 M_1^1 \otimes M_1^1
\end{equation}
In the special case that the vector \mbf{Z} has a complex-valued
normal distribution, then the fourth-order cumulant $C_2^2$ is equal
to zero, such that
\begin{equation}
\label{eqn:normal}
M_2^2 = 2 M_1^1 \otimes M_1^1
\end{equation}
Now, if the components of $\mbf{Z}$ are uncorrelated and have unit
variance, then the second moment $M_1^1$ is the identity matrix; i.e.
\[
\left\{M_1^1\right\}_i^j = \delta_i^j
\]
where $\delta_i^j$ is the Kronecker delta.  Using index notation, the
fourth moment is defined as
\begin{equation}
\left\{M_2^2\right\}^{im}_{jn} = \langle Z_i Z_m Z_j^* Z_n^* \rangle
\end{equation}
However, \Eqn{normal} above says
\begin{equation}
\left\{M_2^2\right\}^{im}_{jn} = {2 M_1^1 \otimes M_1^1}^{im}_{jn} = 2 \delta^i_j \delta^m_n,
\end{equation}
which breaks down when $m=j=a$ and $n=i=b$ and $a \neq b$, such that
\begin{equation}
\left\{M_2^2\right\}^{ab}_{ba} = \langle Z_a Z_b Z_b^* Z_a^* \rangle = \langle |Z_a|^2 |Z_b|^2 \rangle
\end{equation}
where $w w^* = |w|^2$ and $a$ and $b$ are constants; i.e. no summation
over indeces is implied.
Now $|Z_a|^2 \ge 0$ and $|Z_b|^2 \ge 0$; therefore, the expectation of
their product is greater than zero.  However,
\Eqn{normal} dictates that this fourth moment must be zero;
i.e.
\[
\left\{M_2^2\right\}^{ab}_{ba} = 2 \delta^a_b \delta^b_a = 0
\]

%
%

\hrulefill
\subsection{Double contraction of Cardoso (1991)}
\label{sec:car91eq6}

\noindent
In {\it Super-symmetric decomposition of the fourth-order cumulant tensor.
Blind identification of more sources than sensors.},
Cardoso (1991) studies the higher-order statistics of
multivariate complex random values using tensor notation.
Equation (6) of Cardoso (1991) defines two tensor products in terms of
their transformation properties.  The left hand sides of these equations
include a double contraction, the definition of which can be inferred from
the coordinates of the tensors provided in the text that follows, namely
\begin{eqnarray}
\left\{ A\otimes_1B \right\}_{ijkl} & \equiv & a_{ij} b^*_{kl} \\
\left\{ A\otimes_2B \right\}_{ijkl} & \equiv & a_{ik} b^*_{jl}.
\end{eqnarray}
To arrive at the above equations, the double contraction must be defined as
\begin{equation}
\left\{ U M \right\}_{ij}  \equiv u_{ijkl} m_{kl}.
\end{equation}
Note that, in the above equation, no distinction is made between
covariant and contravariant indeces; that is, contractions are
performed over pairs of similar indeces.
To show that the above definition is necessary, start with the
definitions of the outer products,
\begin{eqnarray}
\left( A\otimes_1B \right) M & \equiv & A \tr{ M B^\mathrm{H} } 
\label{eqn:first} \\
\left( A\otimes_2B \right) M & \equiv & A M B^\mathrm{H}.
\label{eqn:second}
\end{eqnarray}
In index notation, the left hand side of \Eqn{first} is
\begin{equation}
\left\{  \left( A\otimes_1B \right) M \right\}_{ij}
  \equiv \left\{ A\otimes_1B \right\}_{ijkl} m_{kl} = a_{ij} b^*_{kl} m_{kl}
\end{equation}
and the right hand side is
\begin{equation}
\left\{ A \tr{ M B^\mathrm{H} } \right\}_{ij}
  = a_{ij} \left\{ M B^\mathrm{H} \right\}_{kk} = a_{ij} b^*_{kl} m_{kl}
\end{equation}
Similarly, the left hand side of \Eqn{second} is
\begin{equation}
\left\{  \left( A\otimes_2B \right) M \right\}_{ij}
  \equiv \left\{ A\otimes_2B \right\}_{ijkl} m_{kl} = a_{ik} b^*_{jl} m_{kl}
\end{equation}
and the right hand side is
\begin{equation}
\left\{ A M B^\mathrm{H} \right\}_{ij}
  = a_{ik} \left\{ M B^\mathrm{H} \right\}_{kj} = a_{ik} b^*_{jl} m_{kl}
\end{equation}

%
%
\hrulefill
\subsection{\Eqns{indexOuterBilinear}{and}{indexSpinorBilinear}
  are consistent with \Eqns{outerBilinear}{and}{spinorBilinear},
  respectively}
\label{sec:eq15to18}

\noindent
Substitute the left hand side of \Eqn{outerBilinear} into the left
hand side of \Eqn{double_contraction} and expand the right hand side
of \Eqn{double_contraction} using the right hand side of
\Eqn{indexOuterBilinear} to arrive at
\begin{equation}
\left\{ \left( \outerBilinear{\bf A}{\bf B} \right) \,\mbf{:}\, \mbf{\rho} \right\}_i^j
  = A_i^j B_k^l \rho_l^k.
\end{equation}
Now expand the right hand side of \Eqn{outerBilinear} using
index notation to arrive at the same result
\begin{equation}
\left\{ {\bf A}\tr{\mbf{\rho}{\bf B}} \right\}_i^j
  = A_i^j \left\{ \mbf{\rho}{\bf B} \right\}_l^l
  = A_i^j B_k^l \rho_l^k
\end{equation}

Similarly, substitute the left hand side of \Eqn{spinorBilinear} into the left hand side
of \Eqn{double_contraction} and expand the right hand side of \Eqn{double_contraction} using
the right hand side of \Eqn{indexSpinorBilinear} to arrive at
\begin{equation}
\left\{ \left( \spinorBilinear{\bf A}{\bf B} \right) \,\mbf{:}\, \mbf{\rho} \right\}_i^j
  = A_i^l B_k^j \rho_l^k.
\end{equation}
Now expand the right hand side of \Eqn{spinorBilinear} using index notation to
arrive at the same result
\begin{equation}
\left\{ {\bf A}\mbf{\rho}{\bf B} \right\}_i^j
  = A_i^l \left\{ \mbf{\rho}{\bf B} \right\}_l^j
  = A_i^l B_k^j \rho_l^k
\end{equation}

%
%
%
\hrulefill
\subsection{Verification of \Eqn{Mueller_to_tensor}}

\noindent
\Eqn{Mueller_to_tensor} can be verified by substituting it into the
right hand side of \Eqn{tensor_to_Mueller}, yielding
\[
\mathrm{RHS}=
\frac{1}{2} \dc{\pauli{\irow}}{\dc{\left( \frac{1}{2} M_\jrow^\jcol \outerBilinear{\pauli{\jrow}}{\pauli{\jcol}} \right)}{\pauli{\icol}}}
=
\frac{1}{4} M_\jrow^\jcol \dc{\pauli{\irow}}{\dc{\left( \outerBilinear{\pauli{\jrow}}{\pauli{\jcol}} \right)}{\pauli{\icol}}}
\]
The above expression can be reorganized using \Eqn{outerBilinear} to yield
\[
\mathrm{RHS}=
\frac{1}{4} M_\jrow^\jcol \left(\dc{\pauli{\irow}}{\pauli{\jrow}}\right)\left(\dc{\pauli{\jcol}}{\pauli{\icol}} \right).
\]
Then, use
\[
\pauli{\irow}\,\mbf{:}\,\pauli{\icol}=\tr{\pauli{\irow}\pauli{\icol}}=2\delta_{\irow\icol}.
\]
to arrive at the left hand side of \Eqn{tensor_to_Mueller} - done!
\[
\mathrm{RHS}=
M_\jrow^\jcol \delta_{\irow\jrow} \delta_{\icol\jcol} = M_\irow^\icol = \mathrm{LHS}.
\]

%
%
%

\hrulefill
\subsection{Derivation of \Eqn{Stokes_cumulant}}

\noindent
Starting with ${\bf U} = \mbf{\kappa}_{2;2}(\mbf{e})$,
substitute \Eqn{fourth_complex_cumulant} into \Eqn{tensor_to_Mueller},
multiply by two, and apply \Eqns{outerMueller}{and}{spinorMueller} to arrive at
\begin{eqnarray}
Q_\irow^\icol &=& 2 M_\irow^\icol =
\pauli{\irow}\mbf{:}\,\mbf{\kappa}_{2;2}(\mbf{e})\mbf{:}\,\pauli{\icol} 
\nonumber \\
&=& \pauli{\irow}\mbf{:}\,\left(
\langle\outerBilinear{\mbf r}{\mbf r}\rangle
- \outerBilinear{\mbf\rho}{\mbf\rho}
- \spinorBilinear{\mbf\rho}{\mbf\rho} \right) \mbf{:}\,\pauli{\icol}  
\nonumber \\
&=& \langle 2\left\{ \outerMueller{\mbf r}{\mbf r} \right\}_\irow^\icol \rangle 
   - 2\left\{ \outerMueller{\mbf \rho}{\mbf \rho} \right\}_\irow^\icol
   - 2\left\{ \spinorMueller{\mbf \rho}{\mbf \rho} \right\}_\irow^\icol.
\end{eqnarray}
Then apply \Eqns{outerStokes}{and}{spinorStokes} to express the above
equation in terms of the instantaneous and population mean Stokes
parameters, $s$ and $S$, associated with $\mbf r$ and $\mbf\rho$,
respectively,
\begin{equation}
{\bf Q}
= \langle s \otimes s^T \rangle - S \otimes S^T - \spinorBilinear{S}{S}
=  {\bf C} -\spinorBilinear{S}{S}
\end{equation}
where the last equality follows from \Eqn{Lorentz_transformation} and the definition of
{\bf C} following \Eqn{Stokes_cumulant}.

%
%
%
\hrulefill
\subsection{Noise bias is consistent with Equation (9) of Cordes \& Hankins (1977)}

\noindent
Assuming that the noise is normally distributed and unpolarized, the last
two terms of \Eqn{covariance-noise} are
\begin{equation}
{\bf B} = {\bf C}_\mathrm{N} + \outerSymm{S_\mathrm{S}}{S_\mathrm{N}} = \spinorBilinear{S_\mathrm{N}}{S_\mathrm{N}} + \outerSymm{S_\mathrm{S}}{S_\mathrm{N}}
\end{equation}
and the biases to the variances of the Stokes parameters are given by
\begin{equation}
  b_\irow = B_{\irow\irow} = S_\mathrm{N,\irow} \left( S_\mathrm{N,\irow} + 2 S_\mathrm{S,\irow} \right)
             - \frac{1}{2}\eta_{\irow\irow} S_\mathrm{N,0} \left( S_\mathrm{N,0} + 2 S_\mathrm{S,0} \right)
\end{equation}
Noting that $S_\mathrm{N,j}=0$ for $j>0$, the biases to the variances of all four
Stokes parameters are equal to
\[
b_\irow = \frac{1}{2} S_\mathrm{N,0} \left( S_\mathrm{N,0} + 2 S_\mathrm{S,0} \right)
\]
When comparing the above bias estimate with Equation (9) of Cordes \&
Hankins (1977), note that $S_\mathrm{S,0} = \langle I_{r_{on}} \rangle +
\langle I_{l_{on}} \rangle$ and, for unpolarized system and sky noise,
$\langle I_{r_{off}} \rangle = \langle I_{l_{off}} \rangle = S_\mathrm{N,0} /
2$.  The biases to the total intensity and circular polarization are
then
\[
C_I = C_V = \frac{1}{4} S_\mathrm{N,0}
\left[ \left( S_\mathrm{N,0} + 4 \langle I_{r_{on}} \rangle \right)
    +  \left( S_\mathrm{N,0} + 4 \langle I_{l_{on}} \rangle \right)
    \right]
    = \frac{1}{2} S_\mathrm{N,0} \left( S_\mathrm{N,0} + 2 S_\mathrm{S,0} \right).
\]

%
%
%
\hrulefill
\subsection{Derivation of \Eqn{covariance-modulation}}

\noindent
Let the modulated instantaneous Stokes parameters $s^\prime=us$ and the covariances between
them
\begin{eqnarray}
{\bf C}^\prime
  & = & \langle s^\prime \otimes s^\prime \rangle
      - \langle s^\prime \rangle \otimes \langle s^\prime \rangle \\
  & = & \langle u^2 \rangle \langle s \otimes s \rangle
      - \langle u \rangle^2 \langle s \rangle \otimes \langle s \rangle
\end{eqnarray}
where the latter equality arises because $s$ and $u$ are statistically
independent (e.g.\ therefore, $\langle us \rangle = \langle u \rangle
\langle s \rangle$).  Now let $u=\langle u \rangle + \delta u$ and
$s=\langle s \rangle + \delta s$, which yields
\begin{eqnarray}
{\bf C}^\prime
& = & \left(\langle u \rangle^2 + \langle \delta u^2 \rangle\right)
      \left(\langle s \rangle \otimes \langle s \rangle + \langle \delta s \otimes \delta s \rangle \right)
      - \langle u \rangle^2 \langle s \rangle \otimes \langle s \rangle \\
& = &  \langle u \rangle^2 {\bf C} + \varsigma_u^2 \langle s \rangle \otimes \langle s \rangle + \varsigma_u^2 {\bf C} \\
& = & \left(\varsigma_u^2 + 1\right) {\bf C} + \varsigma_u^2 S \otimes S
\end{eqnarray}
where
the first equality uses $\langle \delta u \rangle = 0$ and
    $\langle \delta s \rangle = 0$;
the second equality uses $\langle
\delta u^2 \rangle = \varsigma_u^2$ and $\langle \delta s \otimes
\delta s \rangle = {\bf C}$; and
the last equality follows from
$\langle u \rangle = 1$ and $\langle s \rangle = S$.

%
%
%

\hrulefill
\subsection{Derivation of \Eqn{rectangular_modulation}}

After integration over $n^\prime$ instances of the instantaneous
Stokes parameters, the sub-sample mean total intensity
$\bar{S}_0^\prime$ has a population mean
$S_0\equiv\langle\bar{S}_0^\prime\rangle=1$ and variance
$\bar{\varsigma}_0^{\prime2}=(2n^\prime)^{-1}$.  Each instance of the
sub-sample mean is multiplied by a statistically independent and
lognormally distributed variate $u$ with population mean $\langle u
\rangle = \sqrt{\xi}$ and variance $\varsigma_u^2= \xi^2-\xi$, where
$\xi=\exp(\varsigma^2)$ and $\varsigma$ is the standard deviation of
the normally distributed variate $v$ used to generate $u=\exp(v)$.

The amplitude-modulated sub-sample mean total intensity
$\bar{S}_0^{\prime\prime} \equiv u\bar{S}_0^\prime$ has population
mean
$\langle\bar{S}_0^{\prime\prime}\rangle=\langle u\rangle$,
variance (Goodman 1960)
\begin{equation}
  \bar{\varsigma}_0^{\prime\prime2}
  = S_0^2\varsigma_u^2 + \langle u\rangle^2 \bar{\varsigma}_0^{\prime2}
  + \varsigma_u^2 \bar{\varsigma}_0^{\prime2}
  = \varsigma_u^2 \left(1+\frac{1}{2n^\prime}\right)
  + \langle u\rangle^2\frac{1}{2n^\prime},
\end{equation}
and squared modulation index
\begin{equation}
  \beta^{\prime\prime2}\equiv\frac{\bar{\varsigma}_0^{\prime\prime2}}{\langle \bar{S}_0^{\prime\prime}\rangle^2}
  = \frac{\varsigma_u^2}{\langle u\rangle^2}\left(1+\frac{1}{2n^\prime}\right)
  +\frac{1}{2n^\prime}.
\end{equation}
After integrating over $N^\prime=n/n^\prime$ statistically independent
instances of $\bar{S}_0^{\prime\prime}$, the square of the modulation
index of the sample mean Stokes parameters $\bar{S}$ is given by
\begin{equation}
  \beta^2 = \frac{\beta^{\prime\prime2}}{N} = {1\over n}\left[\left(\xi-1\right)
                    \left(n^\prime+\frac{1}{2}\right)
                    +\frac{1}{2}\right].
\end{equation}

\end{appendix}

\end{document}